\documentclass[
  aps,
  prb,
  reprint,
  superscriptaddress,
  amsfonts,
  amssymb,
  amsmath,
  floats,
  footinbib,
  longbibliography
]{revtex4-2}
\usepackage[pdftex]{graphicx}
\usepackage{subfigure}
\usepackage{dsfont}
\usepackage{lipsum}
\usepackage{dcolumn}
\usepackage{bm}
\usepackage{color}
\usepackage{physics}
\usepackage{multirow}
\usepackage[pdftex,colorlinks=true, linkcolor=blue,citecolor=blue,filecolor=blue]{hyperref}
\usepackage{comment}
\usepackage[bb=boondox]{mathalfa}
\allowdisplaybreaks[4]

\usepackage[normalem]{ulem}

\begin{document}
\title{Electronic correlations and fluctuating lattice distortions in vanadium dioxide}

\author{Antonio Picano} 
\affiliation{Coll\`{e}ge de France, PSL Research University,
11 Place Marcelin Berthelot, 75321 Paris, France}
\author{Martin Eckstein}
\affiliation{Institute of Theoretical Physics, University of Hamburg, 20355 Hamburg, Germany}
\affiliation{The Hamburg Centre for Ultrafast Imaging, Hamburg, Germany}
\author{Francesco Grandi}
\thanks{Work begun at the institutions listed; the author is currently unaffiliated.}
\affiliation{Institute for Theory of Statistical Physics, RWTH Aachen University, 52056 Aachen, Germany}
\affiliation{Institut f\"ur Theoretische Physik und Astrophysik and W\"urzburg-Dresden Cluster of Excellence ct.qmat, Universit\"at W\"urzburg, 97074 W\"urzburg, Germany}

\begin{abstract}

Metal-insulator transitions in correlated materials are often accompanied by a change of crystal structure. They are commonly described within a coherent lattice approximation, which combines a correlated treatment of the electrons with a lattice represented by one or a few classical distortion coordinates fixed by minimizing an energy. The structural degrees of freedom then carry no entropy of their own, even though the partition of the transition entropy between electrons and lattice is often what decides the transition temperature. Here we develop a stochastic semiclassical extension of dynamical mean-field theory in which correlated electrons and fluctuating lattice distortions are evolved together, extending previous formulations from linear to nonlinear electron-phonon interactions and from a single to several coupled lattice modes. The resulting non-conservative Langevin equations have deterministic forces, damping, and correlated noise generated self-consistently by the interacting electronic subsystem. Applying the approach to a minimal two-orbital model containing the two symmetry-distinct distortions of the monoclinic phase of vanadium dioxide, we find that the two distortions melt at well separated temperatures, giving an insulating, an intermediate metallic, and a high-symmetry phase. The two transitions have different origins: the melting of the dimerization is driven by the coupling to the correlated electrons and is captured already within the coherent lattice approximation, whereas the restoration of the undistorted structure requires the entropy of the fluctuating lattice. Since the electronic subsystem is treated within nonequilibrium dynamical mean-field theory, the framework can be extended in future work to photoexcited systems with nonthermal electronic distributions.

\end{abstract}
\maketitle

\section{Introduction}
\label{sec:introduction}

In many strongly correlated materials, a metal-insulator transition is accompanied by a change of crystal structure, and it is a long-standing question whether the transition is driven by the electronic degrees of freedom, by the lattice, or by their cooperation. A powerful strategy has been the coherent lattice approximation, which combines a correlated treatment of the electrons with a description of the lattice in terms of one or a few classical distortion coordinates, fixed by minimizing an energy. In this way, DFT+DMFT calculations have located the structural instability accompanying the Mott transition in V$_2$O$_3$~\cite{Leonov2015_PRB} and identified the coupling between electronic disproportionation and a specific structural mode as the mechanism of the transition in the rare-earth nickelates~\cite{Peil2019_PRB,GeorgescuMillis2022_CommunPhys}. The resulting energy landscape, as a function of the electronic state  and the coherent lattice order parameters, can also be propagated in time, which  has recently been used to follow the photoinduced insulator--metal transition in strained Ca$_2$RuO$_4$~\cite{Verma2024_NatPhys}.

In all of these descriptions the lattice is represented by a single deterministic and spatially homogeneous distortion, which responds to the mean electronic force but does not fluctuate. The structural degrees of freedom therefore contribute no entropy of their own, even though that contribution can be decisive. In vanadium dioxide (VO$_2$), the large vibrational entropy of the rutile phase is a major contribution to its thermodynamic stabilization~\cite{Budai2014_Nature}, and first-principles estimates attribute the transition entropy predominantly to the phonons~\cite{Mellan2019_PRB}, whereas estimates based on electronic transport assign most of it to the electronic subsystem~\cite{Paras2020_PRB}. In other systems the balance can shift decisively toward the electrons: in model descriptions of V$_2$O$_3$, it is the entropy of the local moments of the correlated insulator that stabilizes the paramagnetic insulating phase at finite temperature~\cite{Sandri2013_PRB}. Neither contribution can therefore be assumed to dominate, motivating a  description beyond the coherent lattice approximation in which electronic and structural fluctuations are treated on equal footing.

VO$_2$ is a particularly rich realization of this competition. At ambient pressure it undergoes at $T_c\simeq 340$~K a first-order transition in which an electronic metal-insulator transition occurs concomitantly with a structural transformation from a low-temperature monoclinic M$_1$ phase to the high-temperature rutile (R) phase~\cite{Morin1959,Park2013,Chen2017,delValle2019,Pouget2021_CRP}. The M$_1$ phase is a paramagnetic insulator characterized by two coupled distortions of the V sublattice, a dimerization along the rutile $c_R$ axis and a transverse tilting of the V--V dimers, while the R phase is an undistorted paramagnetic metal~\cite{Goodenough1971_JSolStChem}. Whether the insulating state is driven by the lattice~\cite{Goodenough1971_JSolStChem} or electronic correlations~\cite{Pouget1974_PRB,Zylbersztejn1975_PRB} has long been debated, and recent results indicate that the two mechanisms cooperate rather than compete~\cite{Biermann2005_PRL,Haas2024_PRR}. Experiments further indicate that the electronic response, including strong local correlations and orbital polarization~\cite{Dolmantas2026_PRR}, and the structural distortions need not remain locked throughout the transition: metallization has been reported while the lattice still retains a monoclinic or monoclinic-like structure~\cite{Yao2010_PRL,Laverock2014_PRL,Kim2006_PRL,Laverock2018_PRL}, and a related metal-like monoclinic state has been reported after ultrafast photoexcitation~\cite{Morrison2014_Science}, although its existence and interpretation remain debated~\cite{Vidas2020_PRX,Xu2023_NatCommun,Guo2025_NatCommun}.

That the two subsystems can evolve on distinct scales points to the important role of lattice fluctuations in VO$_2$ \cite{Kitou2024_PRB,delValle2026_PRB}. The role of the lattice dynamics is also evidenced by an oxygen isotope effect on the transition temperature~\cite{Rischau2023}. At equilibrium, spatially resolved infrared measurements reveal nanoscale coexistence of metallic and insulating regions in the vicinity of the first-order transition~\cite{Qazilbash2007_Science}. Out of equilibrium, the relevance of fluctuations becomes even more direct. Ultrafast diffuse X-ray scattering showed that photoexcitation rapidly disorders the V--V dimers rather than simply driving a spatially coherent displacement along the equilibrium structural coordinate~\cite{Wall2018}. Subsequent thermal-diffuse-scattering measurements demonstrated that this increase of lattice disorder cannot be understood from the adiabatic potential-energy surface alone: nonconservative forces generated by electron--lattice collisions substantially accelerate the disordering process~\cite{deLaPenaMunoz2023_NatPhys}. Spatially and spectrally resolved ultrafast X-ray imaging has, in turn, revealed a heterogeneous response at longer times and has shown how the identification of transient phases, including the proposed monoclinic metal, can depend crucially on whether spatial structure is resolved~\cite{Johnson2023_NatPhys}. Recent work has further emphasized the thermodynamic role of the incoherent phonon population in the photoinduced transition~\cite{Bagchi2026_arXiv}. Taken together, these results indicate that fluctuations and disorder are not merely corrections to a homogeneous transition pathway, but they must be described consistently together with the electronic and structural order parameters.

Meeting this requirement is difficult: Fluctuating distortions break translational invariance, so the electronic problem must be solved for an inhomogeneous and time-dependent lattice configuration; at the same time, the electronic state cannot be treated at a single-particle level, since the correlated electrons determine the forces, the damping, and the noise that the lattice itself experiences. It is therefore natural to address this question first in a minimal model that retains the competing ingredients identified above, but is simple enough that correlated electrons and stochastic lattice dynamics can be evolved at the same time. Ref.~\cite{Grandi2020_PRR} introduced a minimal microscopic description for VO$_2$, containing exactly these ingredients: local electronic correlations, multiorbital physics, and the two symmetry-distinct components of the M$_1$ distortion. Solved with dynamical mean-field theory (DMFT)~\cite{Georges1996,Aoki2014} within an adiabatic Born--Oppenheimer description of the lattice, the model can produce a first-order transition between an undistorted metal and a distorted insulator and an intermediate distorted metallic phase that can be associated with the experimentally discussed monoclinic metal. This framework also makes explicit that the electronic and lattice entropies can favor the transition in different ways. To disentangle their roles, however, Ref.~\cite{Grandi2020_PRR} analyzes complementary limiting situations in which the transition is driven predominantly by the electronic entropy or by the lattice entropy. Thus, while electrons and phonons are both essential ingredients of the Hamiltonian, their fluctuating thermal contributions are not evolved simultaneously in a self-consistent dynamical description. Related quantum-classical and tensor-network approaches have recently explored the coupled photoinduced dynamics of simplified VO$_2$ models~\cite{Zhang2024_arXiv,Brahms2024_arXiv}.

A framework for treating precisely this kind of feedback between correlated electrons and fluctuating lattice degrees of freedom was developed in Refs.~\cite{Picano2023_PRB,Picano2023_CDW}. Starting from the Keldysh path integral, the lattice coordinates are treated semiclassically while the electronic problem is solved quantum mechanically. Expanding in the quantum component of the lattice field yields stochastic equations of motion in which the mean electronic force, the damping, and the noise correlations are determined self-consistently by the electronic Green's functions and response functions. When combined with an inhomogeneous DMFT description, this approach gives access to spatially resolved lattice fluctuations and was used to describe the formation of an inhomogeneously disordered state across a photoinduced charge-density-wave transition~\cite{Picano2023_CDW}. The previous formulation is, however, restricted to a single structural coordinate with a linear, Holstein-type electron--phonon coupling, whereas the minimal description of VO$_2$ in Ref.~\cite{Grandi2020_PRR} involves two structural modes, one of which couples nonlinearly to the electronic orbital polarization.

In this work, we extend the stochastic semiclassical formalism to this nonlinear, two-component electron--phonon problem, treating correlated electrons within DMFT and lattice fluctuations self-consistently. The model combines two correlated electronic orbitals with semiclassical dimerization and tilting modes, coupled linearly to the electronic density and quadratically to the orbital polarization, respectively. The semiclassical expansion consequently generates not only self-consistent mean forces and damping but also a matrix of stochastic forces whose amplitudes and cross correlations depend on the instantaneous lattice configuration. In this way, the electronic state determines the stochastic lattice dynamics while the spatially fluctuating distortions continuously reshape the local electronic problem. The resulting theory allows us to address, within a single framework, an insulating phase, an intermediate low-symmetry metal, and a high-symmetry metal, the analogues in our model of the M$_1$ insulator, the monoclinic metal, and the rutile metal of VO$_2$, and to determine how their stability is modified by electronic and lattice fluctuations.

The paper is organized as follows. In Sec.~\ref{sec:model} we introduce the model. In Sec.~\ref{sec:semiclassical} we summarize the stochastic semiclassical equations and describe the electronic solution. In Sec.~\ref{sec:results} we present the results: the equilibrium phase diagram, the microscopic picture of the two structural transitions, the corresponding electronic spectra, and an assessment of which ingredients of the stochastic semiclassical approach are essential beyond an adiabatic Born--Oppenheimer description. Finally, in Sec.~\ref{sec:conclusions} we summarize our conclusions.


\section{The model}
\label{sec:model}

We consider a minimal model which captures the interplay between electronic correlations and the main structural distortions involved in the metal-insulator transition of VO$_2$ \cite{Grandi2020_PRR,Zhang2024_arXiv,Brahms2024_arXiv}. The model contains two electronic bands, and two local lattice coordinates $X_1$ and $X_2$ in each unit cell. The electronic interaction favors Mott localization, while the two lattice modes mimic the dimerization and tilting distortions of VO$_2$, respectively. The dimerization couples to the local charge and drives a charge-density-wave instability, whereas the tilting acts as a crystal field and favors an orbital polarization. Our goal is not a microscopic description of VO$_2$, but rather to capture the interplay of these three effects while retaining both local electronic correlations and lattice fluctuations. We therefore study the model on a Bethe lattice with infinite coordination number $Z$ (Fig.~\ref{fig:dimer_bethe}), which allows for considerable simplifications within the inhomogeneous DMFT framework for the treatment of the lattice fluctuations.

\subsection{Local Hamiltonian}

The local electronic and lattice degrees of freedom are described by the on-site Hamiltonian
\begin{align}
    \label{eq:ham_loc}
    H_j = H_U 
    -\sqrt{2\Omega}gX_{1,j}O_{j,1}
    -\frac{\Omega\Delta}{2}X_{2,j}^2O_{j,2}
    +H_{\mathrm{ph},j},
\end{align}
where $H_U=\frac{U}{2} n_j (n_j-1)$ is the local Hubbard repulsion, with  $n_j=\sum_{a,\sigma}n_{j,a,\sigma}$ for the two bands $a=1,2$. We work at quarter filling, $\langle n_j\rangle=1$, with one electron per site in the two-band model. The dimerization mode $X_1$ couples linearly to the density in band $1$,
\begin{align}
    \label{eq:dim}
    O_{j,1}=\sum_\sigma c^\dagger_{j,1,\sigma}c_{j,1,\sigma}-1.
\end{align}
A staggered $X_1$ therefore produces an alternation of the local charge in band $1$, mimicking the  dimerization of the vanadium atoms, and can  open a gap in band $1$. The second mode, $X_2$, describes the tilting distortion. It couples quadratically to the orbital polarization,
\begin{align}
    \label{eq:cf}
    O_{j,2}=n_{j,1}-n_{j,2},
\end{align}
and hence acts as a crystal field that lifts the degeneracy between the two bands. Because the coupling is quadratic in $X_2$, the electronic energy is independent of the sign of the tilting. A finite $X_2$ can thus transfer charge from one band to the other; once band $2$ is depleted, the dimerization can open a charge gap in the remaining occupied band \cite{Goodenough1971_JSolStChem}, such that the two structural distortions cooperatively stabilize the insulating phase.

The dynamics of the two lattice coordinates is determined by the Hamiltonian 
\begin{align}
\label{eq:ham_ph1}
H_{\mathrm{ph},j}
=\frac{1}{2}(P_{1,j}^2+P_{2,j}^2)
+\mathcal{V}(X_{1,j},X_{2,j}),
\end{align}
with the phenomenological potential  for improper ferroelectrics \cite{Kumar2017_PRB},
\begin{align} \label{eq:ham_ph}
	 \mathcal{V}& (X_{1,j}, X_{2,j}) = \frac{\Omega^2}{2} ( X_{1,j}^2 + X_{2,j}^2 )  + \frac{\mu_1\Omega^2 }{4} (2 X_{1,j} X_{2,j})^2 \nonumber \\
	 &+ \frac{\mu_2 \Omega^2}{4} (X_{1,j}^2 - X_{2,j}^2)^2  
	+ \frac{\nu \Omega^3}{6}  (X_{1,j}^2 + X_{2,j}^2)^3.
\end{align}
Here $P_{1,j}$ and $P_{2,j}$ are the conjugate momenta and $\Omega$ is the bare phonon frequency. For $\mu_1\neq\mu_2$, the lattice potential has a discrete C$_4$ symmetry. Together with the different electronic couplings of $X_1$ and $X_2$, Eq.~\eqref{eq:ham_loc}, the symmetry of the full problem is reduced to $\mathbb{Z}_2\times\mathbb{Z}_2$.

\subsection{Lattice Hamiltonian}

\begin{figure}
    \centerline{\includegraphics[width=0.45\textwidth]{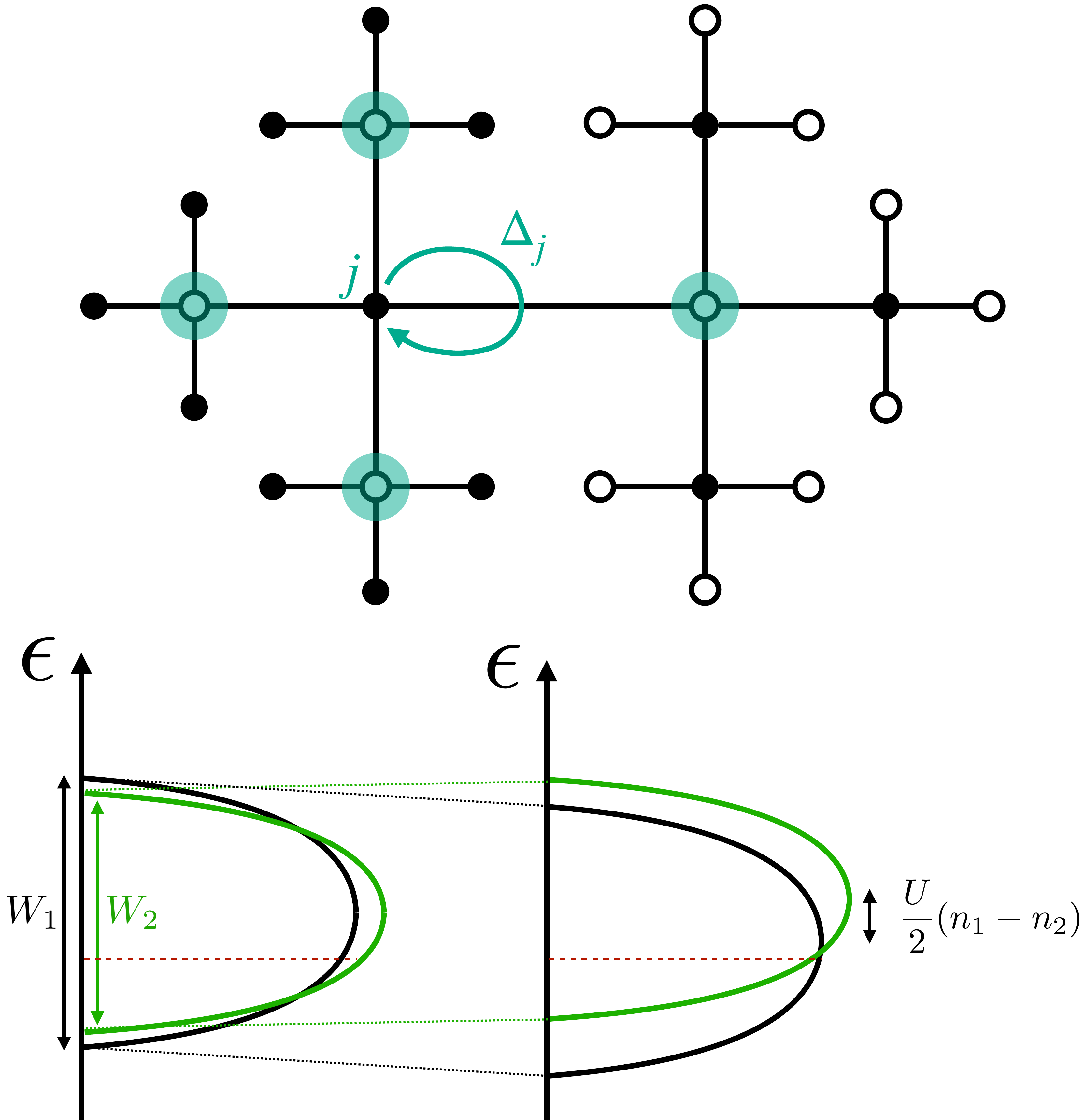}}
	\caption{Upper panel: A portion of the Bethe lattice with coordination number $Z=4$, with the two sublattices A and B shown as open and filled circles. Within DMFT, the environment of a site $j$ is replaced by a self-consistent bath $\Delta_j$, Eq.~\eqref{self_cons}, which is determined by the Green's functions on the opposite sublattice. Lower panel: On the left, we show a sketch of the spectral functions in the undistorted rutile phase without the Hubbard interaction. On the right, we illustrate the mean-field effect of the Hubbard interaction, i.e., a crystal field splitting contribution. The red dashed lines show the occupation. The dotted black (green) lines show the shift down (up) in energy of band $1$ ($2$) due to the Hubbard interaction.}
    \label{fig:dimer_bethe}
\end{figure}

For the hopping on the Bethe lattice, we choose the rescaled intra-orbital nearest-neighbor hoppings $J_{a,a}/\sqrt{Z}$ for orbitals $a=1,2$, and the inter-orbital nearest-neighbor hopping $J_{a,\bar a}/\sqrt{Z}$, where $\bar a=2$ for $a=1$ and vice versa. In the absence of interactions and inter-orbital hopping, the two orbitals correspond to semi-elliptic bands,
\begin{align}
    \label{eq:bethe_dos}
    \mathcal{D}_a(\epsilon)
    =\frac{4}{\pi W_a}
    \sqrt{1-\left(\frac{2\epsilon}{W_a}\right)^2}.
\end{align}
We choose $J_{1,1}=J_0+\delta J/2$ and $J_{2,2}=J_0-\delta J/2$, with $J_0=0.4875$ and $\delta J=0.025$, corresponding to bandwidths $W_1=2$ and $W_2=1.9$, respectively, and take $J_{1,2}=J_{2,1}=J'=0.1$. The slightly larger bandwidth of band $1$ favors its occupation already in the undistorted phase, mimicking the preferential occupation of the a$_{1\mathrm{g}}$ state in rutile VO$_2$, see Fig.~\ref{fig:dimer_bethe} (lower panel). The Hubbard interaction further enhances this occupation imbalance. The inter-orbital hopping allows charge transfer between the two bands and is retained from Ref.~\cite{Grandi2021_arXiv}. It plays only a marginal role in equilibrium, but is required for the nonequilibrium extension discussed below. 
In addition to the electronic intersite terms, we include  a nearest neighbor  phonon coupling
\begin{align}
    \label{ph-inte}
    H_{\mathrm{ph}}^{\mathrm{inter}}
    = \frac{2 J_{\mathrm{ph}}\Omega}{Z} \sum_{\langle i,j\rangle}
    \left(X_{1,i}X_{1,j}-X_{2,i}X_{2,j}\right)
\end{align}
with $J_{\mathrm{ph}}>0$,  which favors a staggered dimerization and a uniform tilting.

The  parameters will be fixed to $U=1.5$, $\Omega=0.155$, $\mu_1=1.75\times10^{-3}$, $\mu_2=2\mu_1$, $\nu=6.722\times10^{-4}$, $g=0.55$, and $\Delta=0.34$. Together with the electronic parameters $J_0=0.4875$, $\delta J=0.025$, and $J'=0.1$ introduced above, these parameters are chosen to reproduce the qualitative equilibrium phase behavior.  (Note that electron-phonon coupling parameters differ quantitatively from those of Ref.~\cite{Grandi2020_PRR}.  Since we use the semicircular densities of states \eqref{eq:bethe_dos} rather than the quasi-one-dimensional density of states of Ref.~\cite{Grandi2020_PRR}, which would enhance the Peierls instability, different coupling strengths are required to obtain a comparable competition between the electronic and structural instabilities.) At high temperature, the system is in the undistorted phase, with $X_1=X_2=0$. Because of its  slightly larger bandwidth, band $1$ has a larger occupation than band  $2$, $n_{1,A}=n_{1,B}\gtrsim 0.5$ and $n_{2,A}=n_{2,B}\lesssim 0.5$ on the two sublattices (A,B) of the Bethe lattice. Upon lowering the temperature, the lattice potential favors the simultaneous development of the dimerization and tilting distortions. The tilting $X_2$ enhances the orbital polarization, while the staggered dimerization $X_1$ drives a charge-density-wave instability between the two sublattices. In the strongly distorted limit, one therefore approaches, for one choice of the broken-symmetry state, $n_{1,A}\simeq 2$ and $n_{1,B}\simeq n_{2,A}\simeq n_{2,B}\simeq 0$. The symmetry-related choices $(\pm X_1,\pm X_2)$ give the four equivalent distorted states.  

\subsection{Inhomogeneous DMFT}

In the semiclassical formulation introduced below, the phonon coordinates at each site will be represented by stochastic trajectories $X_{j,a}(t)$, such that the electronic lattice problem becomes inhomogeneous. Treating such a fully inhomogeneous problem is numerically challenging on a general lattice, but becomes particularly simple on the Bethe lattice in the limit $Z\rightarrow\infty$: Within DMFT, each lattice site is mapped onto a quantum impurity problem with action $S_j=S_j^{\mathrm{loc}}+S_j^{\mathrm{hyb}}$, where $S_j^{\mathrm{loc}}$ describes the coupled electronic and lattice degrees of freedom at site $j$, while
\begin{align}
    S_j^{\mathrm{hyb}}
    =
    \sum_{a,\sigma}\int_{\mathcal C}dt\,dt'\,
    c^*_{j,a,\sigma}(t)
    \Delta_{j,a}(t,t')
    c_{j,a,\sigma}(t')
\end{align}
describes the hybridization of the electrons with the self-consistently determined bath. On the inhomogeneous Bethe lattice, the self-consistent hybridization function for orbital $a$ and site $j$ is
\begin{align}
    \label{self_cons}
    \Delta_{j,a}
    =
    \frac{1}{Z}\sum_{i\in\text{NN}(j)}
    \left(
    |J_{a,a}|^2G_{i,a}
    +
    |J_{a,\bar a}|^2G_{i,\bar a}
    \right),
\end{align}
where the sum extends over the nearest neighbors of $j$. Equation~\eqref{self_cons} introduces a charge-transfer channel between the two orbitals on top of an orbital-diagonal hopping, while keeping the local problem orbital diagonal: the off-diagonal hybridizations $\Delta_{j,a\bar a}$ that a genuine two-orbital Bethe lattice would generate are absent by construction. The same strategy is used for the square lattice in Ref.~\cite{Sandri2013_PRB}, where the momentum dependence of the inter-orbital hopping is chosen so that the local single-particle density matrix remains diagonal in the orbital indices. Since the Bethe lattice is bipartite, the nearest neighbors of a site on sublattice $A$ ($B$) all belong to sublattice $B$ ($A$). In the limit $Z\rightarrow\infty$, the sum over neighboring sites in Eq.~\eqref{self_cons} self-averages. As a consequence, the hybridization function is the same for all sites of a given sublattice and is determined by the statistical average of the local Green's functions on the opposite sublattice. The inhomogeneous problem can therefore be represented by an ensemble of stochastic trajectories for each sublattice: the local electronic properties $G_j$ and lattice distortions $X_j$ fluctuate from one trajectory to another, but the trajectories on a given sublattice share the same self-consistently determined hybridization function. This statistical DMFT construction allows to retain the local electronic and lattice fluctuations without explicitly treating a fully inhomogeneous lattice.


\section{Summary of the stochastic semiclassical equations}
\label{sec:semiclassical}

\subsection{Semiclassical stochastic equations}
\label{sec:sca_summary}

Within statistical DMFT, the lattice problem is mapped onto an ensemble of impurity problems $j$, each describing interacting electrons coupled to the two local lattice modes $X_{j,1}$ and $X_{j,2}$. To solve these impurity problems, we use the stochastic semiclassical approximation (SCA) introduced in Ref.~\cite{Picano2023_PRB}.  Since the stochastic equations describe an individual impurity problem, we omit the site index \(j\) throughout this subsection. In the SCA, the quantum lattice coordinates are replaced by stochastic classical trajectories $X^{\rm cl}_a(t)$, while the electronic degrees of freedom are treated quantum mechanically and evolve in the presence of the time-dependent lattice configuration. The dynamics of $X^{\rm cl}_a(t)$ is determined by the lattice potential and by the electronic back-action. A stochastic equation of motion for $X^{\rm cl}$ is obtained by integrating out the electrons and expanding the resulting electronic influence functional to second order in the quantum component of $X_a$. To first order, the electrons generate an instantaneous Hellmann--Feynman force, while at second order electronic fluctuations generate friction and noise, determined by the retarded and Keldysh electronic response functions, respectively. In the adiabatic limit, the  instantaneous force would reduce to the force derived from the Born--Oppenheimer potential-energy surface, while friction and noise describe the nonadiabatic electronic back-action. Similar ingredients were introduced phenomenologically to describe the ultrafast disordering of VO$_2$ in Ref.~\cite{deLaPenaMunoz2023_NatPhys}; in the present approach, both are obtained from the electronic correlation functions, without introducing an effective electronic temperature.

In this section, we state the resulting equations. Details of the derivation are given in App.~\ref{app:derivation}. With the vector notation $\mathbf X^{\rm cl}=(X_1^{\rm cl},X_2^{\rm cl})$, the semiclassical dynamics takes the Langevin form
\begin{align}
    \label{eq:sca_compact}
    \ddot{\mathbf X}^{\rm cl}(t)
    =
    \mathbf F[\mathbf X^{\rm cl},\dot{\mathbf X}^{\rm cl}]
    +
    \boldsymbol{\zeta}(t),
\end{align}
where $\boldsymbol{\zeta}(t)$ is the noise, and  the deterministic force is
\begin{align}
    \label{eq:sca_force}
    \mathbf F
    =
    -\boldsymbol{\nabla}_{\mathbf X^{\rm cl}}\mathcal V(\mathbf X^{\rm cl})
    +\mathbf F^{\rm el}
    -\mathbf D\,\dot{\mathbf X}^{\rm cl}
    +\mathbf F^{\rm inter}.
\end{align}
The four terms describe the bare lattice force due to the local potential \eqref{eq:ham_ph}, the instantaneous electronic force, electronic friction, and the intersite lattice coupling, respectively. From Eq.~\eqref{eq:ham_loc},  the electronic Hellmann--Feynman force is given by
\begin{align}
    \label{eq:sca_el_force}
    F^{\rm el}_a
    =
    v_a\langle\hat O_a\rangle_{\rm cl},
\end{align}
with electron--lattice vertices
\begin{align}
    \label{eq:vertex}
    v_1=\sqrt{2 \Omega}\,g,
    \qquad
    v_2=\Delta \Omega X_2^{\rm cl}.
\end{align}
Here the expectation value $\langle\cdots\rangle_{\rm cl}$ is calculated for the electronic model in the presence of the time-dependent lattice displacement $X^{\rm cl}$ (see below). For  $Z\rightarrow\infty$ limit, the intersite phonon force  $\mathbf F^{\rm inter}$ is treated at the mean-field level,
\begin{align}
    \label{eq:sca_inter_force}
    F^{\rm inter}_a
    =
    -\gamma_a\,2\Omega J^{\rm ph}
    \langle X_a^{\rm cl}\rangle_{\rm nn},
\end{align}
with $\gamma_1=1$ and $\gamma_2=-1$; $\langle X_a^{\rm cl}\rangle_{\rm nn}$ denotes the statistical average of the distortion on the opposite sublattice. The friction and noise are determined by the electronic correlation functions  $\chi_{ab}(t,t')=-2i\langle T_{\mathcal C}\hat O_a(t)\hat O_b(t')\rangle^{\rm con}_{\rm cl}$.   Here we assume that the electronic correlation time $\tau_{\rm el}$ is short compared with the characteristic time scale of the lattice dynamics,
\begin{align}
    \tau_{\rm el}\ll\Delta t\ll\Omega^{-1}.
    \label{eq:timescale_sep}
\end{align}
This allows us to take the white-noise limit, in which the friction force becomes instantaneous,
\begin{align}
    \label{eq:sca_damping}
    D_{ab}(t)
    =
    -v_av_b\,
    {\rm Im}\,
    \left.
    \partial_\omega
    \chi^R_{ab}(\omega,t)
    \right|_{\omega=0},
\end{align}
and the noise force is delta-correlated in time, 
\begin{align}
    \label{eq:sca_noise}
    \langle\zeta_a(t)\zeta_b(t')\rangle
    =
    K_{ab}(t)\delta(t-t'),
\end{align}
with $\langle\zeta_a(t)\rangle=0$ and
\begin{align}
    K_{ab}(t)
    =
    -\frac{v_av_b}{2}\,
    {\rm Im}\,
    \chi^K_{ab}(\omega=0,t).
\end{align}

Compared with the single-mode, linear electron--phonon coupling considered in Ref.~\cite{Picano2023_PRB}, there are two important extensions of the formulation in this manuscript: First, the electronic response couples the two lattice modes, so that both the friction and noise covariance are matrices, allowing for cross-damping and correlated fluctuations. Second, the quadratic coupling of the crystal field splitting \eqref{eq:cf} to the tilting mode gives the configuration-dependent vertex $v_2=\Omega \Delta X_2^{\rm cl}$. Consequently, the electronic friction and noise associated with the tilting mode depend on the instantaneous distortion and vanish for $X_2^{\rm cl}=0$.

For electrons in local thermal equilibrium at temperature $T$, the retarded and Keldysh response functions obey the fluctuation--dissipation relation
\begin{align}
{\rm Im}\,\chi^K_{ab}(\omega\rightarrow0)
=
4T\,{\rm Im}
\left.
\partial_\omega\chi^R_{ab}(\omega)
\right|_{\omega=0},
\end{align}
which implies the Einstein relation between noise and friction,
\begin{align}
K_{ab}
=
2T D_{ab}.
\end{align}
Out of equilibrium, this relation no longer holds, and deviations from it characterize the nonthermal electronic environment seen by the lattice. Such nonthermal electronic states are treated explicitly in our approach by evolving the electronic subsystem self-consistently within nonequilibrium DMFT, using the quantum Boltzmann equation (QBE) formulation of Ref.~\cite{Picano2021}, as described in Sec.~\ref{sec:DMFT_loop}. At each time step, the electronic Green's functions and response functions determine the mean force, friction, and noise entering the lattice dynamics, while the instantaneous lattice configuration feeds back onto the electronic Hamiltonian. The resulting scheme thus provides a self-consistent stochastic evolution of the coupled electronic and lattice degrees of freedom without assuming an effective electronic temperature.

\subsection{Inhomogeneous DMFT setup}
\label{sec:DMFT_loop}

To solve the electronic problem within DMFT, we employ the quantum Boltzmann equation (QBE) formulation introduced in Ref.~\cite{Picano2021}, which determines  the local energy distribution function 
$F_{j,a} ( \omega, t)$ and the local spectral function $A_{j,a} (\omega, t)$ at each time and for each site $j$.  The evolution of   $F_{j,a} ( \omega, t)$ is determined by the equation  $\partial_t F_{j,a} ( \omega, t ) = I_{j,a,\omega} [F]$, with scattering integral
\begin{align}
\label{scatt_int}
	& I_{j,a,\omega} [F] = -i \big [ \Sigma^<_{j,a}( \omega, t )  \nonumber \\
	&+ \Sigma^R_{j,a} (\omega, t) F_{j,a} ( \omega, t )  - F_{j,a} ( \omega, t ) \Sigma^A_{j,a} (\omega, t) \big ].
\end{align}
Here $\Sigma_{j,a}(\omega,t)$ denotes the electronic self-energy, which is local within the DMFT approximation, and diagonal in the orbital basis. 
The self-energy  
\begin{align}
    \label{sigma_int}
    \Sigma_{j,a}
    =
    \Sigma^{U}_{j,a}
    +
    \Sigma^{\text{ph}}_{j,a}
\end{align}
contains the Hubbard contribution $\Sigma^{U}_{j,a}$ and a phenomenological bosonic bath contribution $\Sigma^{\text{ph}}_{j,a}$ describing electronic energy dissipation into modes other than the lattice coordinates $X_a$ treated explicitly. We approximate $\Sigma^{U}_{j,a}$ by second-order iterated perturbation theory, while $\Sigma^{\text{ph}}_{j,a}$ is evaluated from the first-order electron--boson diagram for a reservoir at fixed temperature $T_{\rm bath}$; explicit expressions for the two self-energies are given in Apps.~\ref{app:ipt} and~\ref{app:thermal_bath}, respectively. This dissipative channel is relevant on the $100$~fs--ps time scales considered here and eventually drives the electronic subsystem toward a thermal fixed point if the dynamics is started from non-thermal distribution. Finally, within the QBE formalism, photo-doping could be incorporated through an additional reservoir self-energy~\cite{Picano2021}; such driven dynamics will however be left for future work.

\subsection{Timestepping}
\label{sec:timestepping}

As mentioned already above, for the inhomogeneous DMFT simulation on the Bethe lattice, the DMFT hybridization $\Delta_j$ and the nearest neighbor phonon interaction \eqref{eq:sca_inter_force} self-average, and the respective mean values depend only on the sublattice $\alpha=A,B$. In the simulations, we therefore keep an ensemble of $N$ statistically independent trajectories $j$ representative of each sublattice, where each realization contains the coordinates $\mathbf X_{j}(t)=(X_{1,j}(t),X_{2,j}(t))$ at discrete times $t_n=n\delta t$, as well as the electronic spectra distribution functions $F_{a,j}(\omega,t)$. The time-stepping procedure is summarized as follows:\\

\noindent {(I)}
At fixed time $t=t_n$, given $\mathbf X_{j}(t)$ and $F_{a,j}(\omega,t)$, we self-consistently solve the electronic problem to determine the spectra $A_{j,a}(\omega,t_{n})$ and the self-energy $\Sigma_{j,a}(\omega,t_{n})$.  Starting with a guess for the self-energy and the hybridization functions $\Delta_{j,a}(\omega,t_n)$ (e.g., taking the value from the previous time-step),  this involves the following self-consistent procedure (the time argument $t=t_n$ is omitted from all quantities in the following self-consistent cycle for simplicity):
\\

\noindent (I-1)	   
Compute the Green's function
\begin{align} \label{G_ret} 
    G^R_{j,a}(\omega)= \big[\omega^+- h_{a}(X_j) - \Delta_{j,a}^{R} (\omega) -  \Sigma_{j,a}^{R}(\omega)\big]^{-1}\!\!\!, 
\end{align}
where $\omega^+=\omega+ i0^+ +\mu$, and the on-site energies $h_{a}(X_j)$ are determined by the instantaneous classical displacements through the electron--phonon coupling in Eq.~\eqref{eq:ham_loc}. The spectrum is given by $A_{j,a}(\omega)=-\frac{1}{\pi}\text{Im}\,G^R_{j,a}(\omega)$, and lesser and greater Green's functions follow from the instantaneous distribution function and the spectrum,
\begin{align}
    & G_{j,a}^{ <} (\omega) = 2\pi i  A_{j,a}(\omega) F_{j,a}(\omega), \\
    & G_{j,a}^{ >} (\omega) = - 2\pi i A_{j,a}(\omega) \big [1 -  F_{j,a}(\omega)  \big].
\end{align}

\noindent   (I-2)	
Determine the hybridization function $\Delta_{\alpha,a}$ on sublattices $\alpha=A,B$ 
from the self-consistency condition \eqref{self_cons},
\begin{align}
    \label{self_cons1}
    \Delta_{\alpha,a}
    =
    |J_{a,a}|^2
     \langle G_{l,a}\rangle_{l\in\bar\alpha}
    +
    |J_{a,\bar a}|^2
     \langle G_{l,\bar a}\rangle_{l\in\bar\alpha}
\end{align}
where $\langle\cdots\rangle_{l\in\bar\alpha}$ is the average over trajectories in the ensemble $\bar\alpha=B$ for $\alpha=A$ and vice versa. Set $\Delta_{j,a}=\Delta_{\alpha,a}$ for all $j\in \alpha$.
\\

\noindent (I-3)	
Construct the Weiss Green's function:
\begin{align} \label{gweiss}
    &\mathcal{G}_{j,a}^{ R}(\omega)	= \big[ \,\omega^+  - h_{j,a} - \Delta_{j,a}^{R}(\omega)\big]^{-1} \\&\mathcal{G}_{j,a}^{<,>}(\omega) = \mathcal{G}_{j,a}^{R}(\omega) \Delta_{j,a}^{<,>}(\omega)  \mathcal{G}_{j,a}^A(\omega),
\end{align}
at time $t_n$, and thus obtain the self-energy using the expressions given in Apps.~\ref{app:ipt} and~\ref{app:thermal_bath}. Repeat steps (I-1) to (I-3) until convergence.\\
	  
\noindent {(II)}
From the converged electronic Green's functions $G_{j,a}(\omega,t_n)$, we obtain the response and fluctuation kernel $\chi_{j,ab}(\omega,t_n)$ and thus the friction  $\mathbf D_{j}(t_n)$ [Eq.~\eqref{eq:sca_damping}] and noise $\mathbf K_{j}(t_n)$ [Eq.~\eqref{eq:sca_noise}]. Here $\chi_{j,ab}$ is obtained from $G_{j,a}$ using a conventional bubble approximation; explicit expressions in terms of the local Green's functions
are given in Appendix~\ref{app:susceptibilities}. \\

\noindent {(III)}
We propagate the coordinates according to the Langevin  equation \eqref{eq:sca_compact}. Explicitly,
\begin{align}
    \mathbf X_{j}(t_{n+1}) &= \mathbf X_{j}(t_n) + \delta t\,\mathbf P_{j}(t_n),
    \\
    \mathbf P_{j}(t_{n+1}) &= \mathbf P_{j}(t_n) + \delta t\, \mathbf F[\mathbf X_{j}(t_n),\mathbf P_{j}(t_n)] + \mathbf W_j,
\end{align}
where the stochastic increments $W_{j,a}$ are gaussian distributed with mean zero and variance $\langle W_{j,a} W_{j,b}\rangle = \delta tK_{j,ab}(t)$. \\

\noindent {(IV)} The electronic distribution function $F_{j,a}(\omega,t)$ is propagated from $t_n$ to $t_{n+1}$ with an exponential-Euler step. We write the QBE [Eq.~\eqref{scatt_int}], at fixed $\omega$ as
$\partial_t F = Y - 2\Gamma F$, with $\Gamma_{j,a}(\omega) = -\mathrm{Im}\,[\Sigma^R+\Delta^R]_{j,a}(\omega)$ and $Y_{j,a}(\omega) = \mathrm{Im}\,[\Sigma^<+\Delta^<]_{j,a}(\omega)$ evaluated from the self-energies converged at $t_{n+1}$ for the updated lattice configuration and the distribution $F(t_n)$. We then set
\begin{equation}
    F(t_{n+1}) = F(t_n) + \frac{1-e^{-2\Gamma h}}{2\Gamma}\,
    \big[Y - 2\Gamma F(t_n)\big] .
\end{equation}
The relaxation term is thus integrated exactly, which keeps the step stable for arbitrarily large $\Gamma$ (inside the gap, at the zero crossings of $X_1$), while for $\Gamma h \ll 1$ the update reduces to an explicit Euler step.

\begin{figure}
    \centering
    \includegraphics[width=\linewidth]{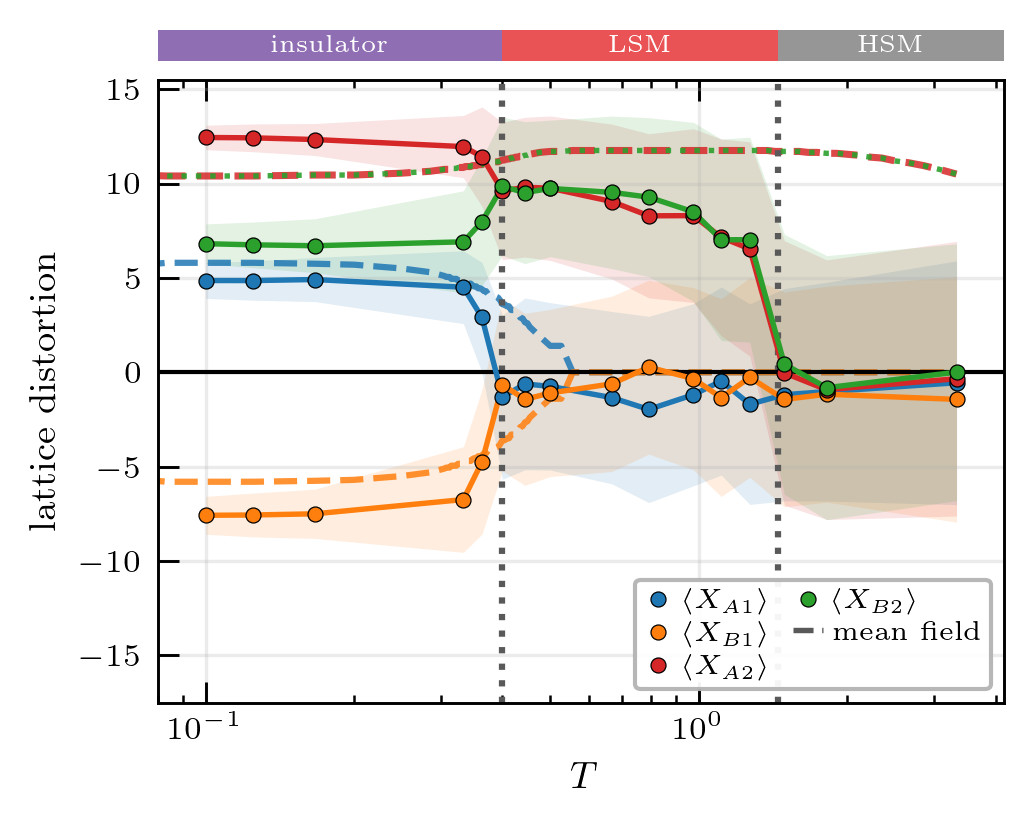}
    \caption{\textbf{Equilibrium phase diagram of the stochastic semiclassical problem.} Ensemble averages over $64$ trajectories, taken over the equilibrated window $t \ge t_\text{eq}$, as a function of temperature $T$. The bar on top marks the three regimes identified in the text. Raw sublattice distortions $\langle X_{A1} \rangle$, $\langle X_{B1} \rangle$, $\langle X_{A2} \rangle$, $\langle X_{B2} \rangle$ (symbols and full lines) are compared with the coherent-lattice approximation (dashed lines) of App.~\ref{app:meanfield}. The  coherent-lattice approximation captures the low-temperature insulating phase and the onset of the low-symmetry metallic regime, but remains distorted at high temperature, highlighting the importance of fluctuations beyond the coherent-lattice approximation in the present parameter regime. The shaded bands indicate the standard deviation of the ensemble (full distribution functions are analyzed in Sec.~\ref{sec:results_landscape}).
}
 \label{fig:equilibriumphasediagram}
\end{figure}

\begin{figure*}
    \centering
    \includegraphics[width=\linewidth]{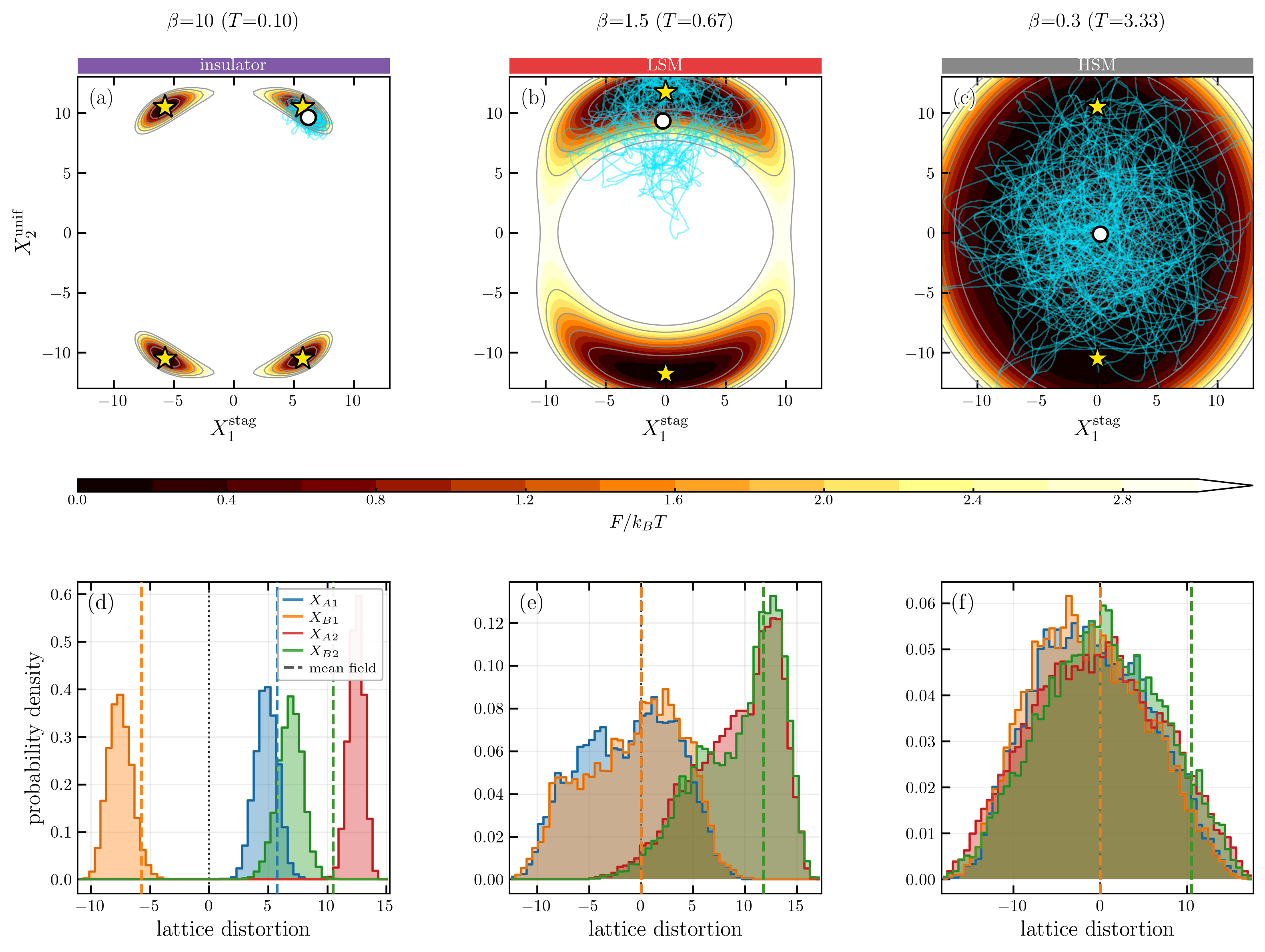}
    \caption{\textbf{Stochastic lattice fluctuations in the mean-field free-energy landscape.}
    Columns correspond to the representative temperatures $\beta=10$ (insulator), $\beta=1.5$ (low-symmetry metal), and $\beta=0.3$ (high-symmetry metal). Top row: mean-field free-energy landscape $F_{\rm MF}(X_1,X_2)/k_BT$ of Eq.~\eqref{eq:F_MF}, with the mean-field minima marked by stars. The stochastic semiclassical trajectories for $t \ge t_{\rm eq}$ are projected onto the same coordinates and superimposed in blue; the white circle denotes the ensemble average. Twenty-four of the 64 trajectories are shown; averages and histograms use the full ensemble. Since the mean-field landscape is defined within the restricted ansatz, the trajectories are projected onto the staggered dimerization $X_1^{\rm stag}=(X_{A1}-X_{B1})/2$, and the uniform tilting $X_2^{\rm unif}=(X_{A2}+X_{B2})/2.$ Bottom row: probability distributions of the four microscopic lattice coordinates accumulated over all trajectories for $t\ge t_{\rm eq}$; the dashed lines indicate the corresponding mean-field values. In the I phase the trajectories remain confined within a single symmetry-broken minimum, with well-separated dimerization distributions and finite tilting distortions. In the low-symmetry metallic phase the dimerization distributions merge around zero while the tilting coordinates remain finite. In the high-symmetry phase the stochastic distributions become centered around the undistorted configuration, whereas the corresponding mean-field minimum remains at finite tilting distortion.}
    \label{fig:MFSemiclEnergies}
\end{figure*}

\section{Results and discussion}
\label{sec:results}

For each temperature $T$ of the electronic bath, we solve the coupled stochastic equations of motion for two ensembles, each containing $64$ statistically independent trajectories, representing the lattice coordinates $\mathbf{X}_{j}(t)=(X_{1,j}(t),X_{2,j}(t))$ and the local electronic spectral and distribution functions on the two sublattices.

The model parameters are $U=1.5$, $J_0=0.4875$, $\delta J=0.025$, and $J'=0.1$ for the electronic sector, with bandwidths $W_1=2$ and $W_2=1.9$, and $\Omega=0.155$, $g=0.55$, $\Delta=0.34$, $\mu_1=1.75\times10^{-3}$, $\mu_2=2\mu_1$, $\nu=6.722\times10^{-4}$, and $J_{\rm ph}=0.1$ for the lattice sector. Identifying the bandwidth $W_1=2$ with the ${\approx}2.6\,$eV width of the vanadium $t_{2g}$ manifold of VO$_2$~\cite{Eyert2002} sets the energy unit $\epsilon_0=W_1/2=1$ to approximately $1.3\,$eV. Accordingly, the Hubbard interaction and phonon frequency correspond to $U\simeq1.95\,$eV and $\Omega\simeq0.20\,$eV, respectively. The bosonic bath is held at the system temperature $T$ [App.~\ref{app:thermal_bath}], with coupling $g_{\rm ph}=0.34$ and Ohmic cutoff $\omega_{\rm ph}=0.2$, corresponding to approximately 0.44 eV and 0.26 eV, respectively, while the lattice Langevin bath has friction $\gamma_{\rm ph}=0.2$, corresponding to 0.26 eV. All simulations use $N=64$ trajectories per sublattice, a time step $\delta t=0.1$, and a frequency grid with spacing $d\omega=0.004$ and cutoff $|\omega|\leq50$. The electronic self-consistency at each time step is iterated to an accuracy of $10^{-6}$.

As an initial state, we choose for every trajectory a fixed reference configuration of the lattice coordinates with vanishing momenta. In the ordered regime, we start from a symmetry-broken configuration with staggered dimerization and uniform tilting, $X_{A1}=-X_{B1}$ and $X_{A2}=X_{B2}$, thereby selecting one of the equivalent broken-symmetry sectors. At the highest temperatures, we instead start from the undistorted configuration, $X=P=0$. The electronic sector is initialized in the self-consistent equilibrium state corresponding to the chosen lattice configuration, with $F_{j,a}(\omega)$ given by the Fermi function at the bath temperature.

Starting from these configurations, we propagate the coupled stochastic dynamics until the ensemble reaches a stationary state. We monitor equilibration through both the ensemble-averaged lattice distortions and their trajectory-to-trajectory variances. As shown in App.~\ref{app:equilibration}, after an initial transient these quantities fluctuate around stationary values at representative temperatures in all three phases. We therefore choose $t_{\rm eq}=100$, safely within this stationary regime, and evaluate equilibrium observables using only data for $t\geq t_{\rm eq}$. Away from the transition regions, simulations initialized from different thermal histories converge to the same stationary state, while the symmetry-breaking seed only selects one of the degenerate ordered sectors. Close to the transitions, equilibration becomes slower because of the enhanced lattice fluctuations discussed in Sec.~\ref{sec:results_phasediagram}.

In the following, we analyze the stationary ensemble to determine the temperature-dependent phase diagram and characterize the structural and electronic properties of the corresponding equilibrium states.

\subsection{Equilibrium phase diagram}
\label{sec:results_phasediagram}

The equilibrium phase diagram obtained from the stochastic dynamics is summarized in Fig.~\ref{fig:equilibriumphasediagram}. We characterize the structural state by the staggered dimerization, $X_1^{\rm stag}=(\langle X_{A1}\rangle-\langle X_{B1}\rangle)/2$, and the uniform tilting, $X_2^{\rm unif}=(\langle X_{A2}\rangle+\langle X_{B2}\rangle)/2$. Upon increasing temperature, these two order parameters disappear at clearly separated temperature scales. At low temperature, both are finite. The dimerization then vanishes at $T=0.4$, while the tilting remains finite up to substantially higher temperatures and disappears only around $T=1.45$. The model therefore exhibits a broad intermediate regime with finite tilting but vanishing dimerization. Within our model description, the three phases we observe share some analogies with the M$_1$ insulator (I), the monoclinic metal \cite{Laverock2014_PRL,Laverock2018_PRL} (LSM), and the rutile metal (HSM) in VO$_2$.

To understand the different origin of the two transitions, it is useful to first consider the lattice sector alone. The local potential $\mathcal V(X_1,X_2)$ has its minimum at $X_1=X_2=0$. However, the intersite lattice interaction \eqref{ph-inte} favors precisely the ordering pattern observed at low temperature: staggered $X_1$ and uniform $X_2$. For the restricted configuration $X_{1,A}=-X_{1,B}\equiv X_1$ and $X_{2,A}=X_{2,B}\equiv X_2$, the pure lattice energy consistent with Eq.~\eqref{eq:sca_inter_force} is
\begin{align}
    E_{\rm latt}(X_1,X_2)
    =
    \mathcal V(X_1,X_2)
    -\Omega J_{\rm ph}(X_1^2+X_2^2).
    \label{eq:pure_lattice_energy}
\end{align}
For the parameters used here, its zero-temperature minima occur at $|X_1|=|X_2| \simeq 4.75$. Thus, already without electrons, the inter-site coupling produces a structural instability in which the two distortions occur together, although their magnitude is cooperatively enhanced by the coupling to the electrons. Importantly, due to the $C_4$ symmetry of the pure lattice energy \eqref{eq:pure_lattice_energy}, the two order parameters are symmetry-related and therefore cannot melt at different temperatures in the pure lattice problem. The transition to the intermediate LSM phase (which will be shown below to be metallic) must therefore originate from the coupling to the electronic degrees of freedom.

This interpretation is supported by comparison with the coherent-lattice approximation described in App.~\ref{app:meanfield}. This approximation retains the self-consistent electronic contribution at the Hartree level but neglects spatial lattice fluctuations and their associated entropy. It corresponds to the stationary limit of the semiclassical equations of motion without lattice fluctuations, with $X_{1,B}=-X_{1,A}$ and $X_{2,B}=X_{2,A}$. Remarkably, the coherent-lattice approximation already reproduces the first transition qualitatively well: the dimerization disappears while the tilting remains finite, at a transition temperature close to that of the stochastic calculation. The emergence of the intermediate regime can therefore be understood as an electronically driven modification of the lattice free-energy landscape, which selectively destabilizes the dimerization while leaving the tilting ordered. In spite of the electronic nature of the intermediate phase, lattice fluctuations nevertheless quantitatively renormalize this transition: within the coherent-lattice approximation the dimerization vanishes at $T\approx0.54$, while in the full stochastic dynamics the I to LSM transition occurs at $T\approx0.36$--$0.40$, i.e., including the lattice fluctuations within the SCA reduces the transition temperature by a factor of approximately $0.7$.

The situation is qualitatively different for the higher-temperature transition. The restricted mean-field solution retains a finite tilting distortion throughout the temperature range considered and therefore never reaches the fully symmetric phase. In the stochastic calculation, by contrast, $X_2^{\rm unif}$ eventually vanishes and the symmetry of the undistorted state is restored. This second transition therefore cannot be explained by the electronic contribution to the mean-field free energy alone. Rather, it reflects the entropy associated with the fluctuating lattice degrees of freedom, which becomes sufficiently important at high temperature to destabilize the remaining tilted state.

\begin{figure*}
    \centering
    \includegraphics[width=\linewidth]{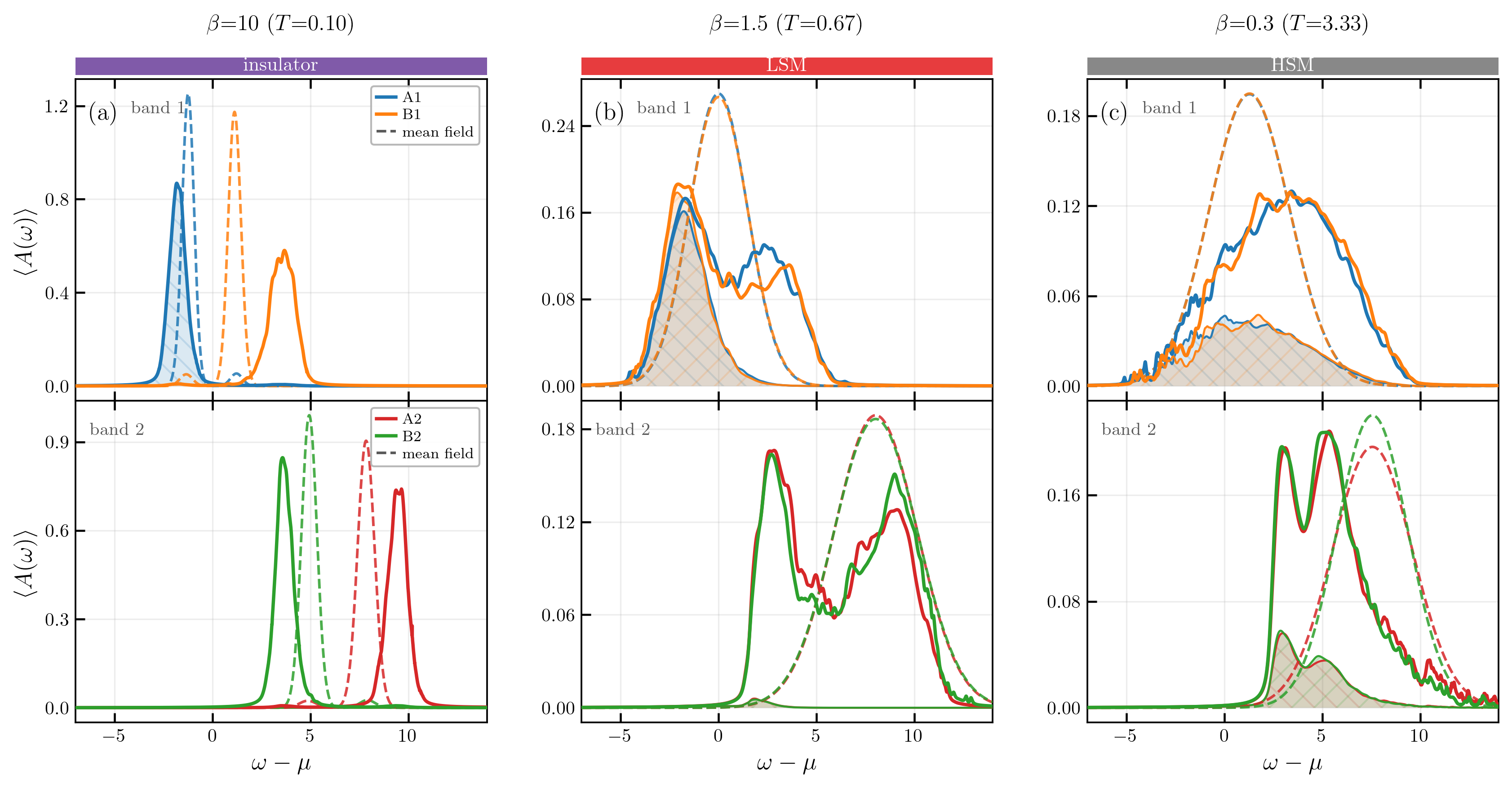}
    \caption{\textbf{Electronic spectral functions across the three equilibrium phases.}
    Columns correspond to the representative temperatures $\beta=10$ (insulator), $\beta=1.5$ (low-symmetry metal), and $\beta=0.3$ (high-symmetry metal). The upper and lower panels show the ensemble-averaged spectral functions $A(\omega)=-\mathrm{Im}\,G^R(\omega)/\pi$ for bands 1 and 2, respectively, with the two sublattices shown by different colors. The shaded regions indicate the occupied spectral weight $A(\omega)F(\omega)$, while the dashed curves denote the corresponding mean-field spectra. In the I phase the dimerization splits the spectrum of band 1 and opens an insulating gap, while the finite tilting distortion shifts band 2 to higher energies. In the LSM phase the dimerization-induced gap collapses whereas the finite tilt maintains a sizable orbital splitting. Finally, in the stochastic HSM phase the disappearance of the tilting distortion leads to strongly overlapping spectra of the two bands, whereas the mean-field solution retains a finite orbital splitting associated with its residual tilting distortion.}
    \label{fig:Semiclassical}
\end{figure*}

\subsection{Microscopic picture of the structural transition}
\label{sec:results_landscape}

While Fig.~\ref{fig:equilibriumphasediagram} identifies the three equilibrium phases through their average lattice distortions, the stochastic semiclassical framework also provides direct access to the underlying equilibrium probability distribution of the lattice degrees of freedom. Figure~\ref{fig:MFSemiclEnergies} illustrates this microscopic picture by combining the stochastic trajectories with the corresponding probability distributions. To facilitate comparison with the free-energy landscape obtained in the coherent lattice approximation (App.~\ref{app:meanfield}), the stochastic trajectories are projected onto the two collective coordinates $X_1^\mathrm{stag}=(X_{A1}-X_{B1})/2$ and $X_2^\mathrm{unif}=(X_{A2}+X_{B2})/2$, which correspond to the staggered dimerization and the uniform tilting distortion, respectively, and which are retained in the coherent lattice approximation.

Deep in the insulating phase ($\beta=10$), the trajectories remain localized close to one of the symmetry-related minima of the coherent-lattice free-energy landscape. The corresponding probability distributions are sharply localized around their symmetry-broken values, reflecting the coexistence of staggered dimerization and finite tilting characteristic of the low-temperature phase. A qualitatively different behavior emerges in the intermediate LSM regime ($\beta=1.5$), where the trajectories are no longer confined along the dimerization direction and fluctuate around $X_1^{\rm stag}=0$, while remaining localized around a finite value of $X_2^{\rm unif}$. The sublattice symmetry breaking associated with the dimerization is therefore lost, such that both the expectation values shown in Fig.~\ref{fig:equilibriumphasediagram} and the distribution functions become identical on the two sublattices within statistical fluctuations, $\langle X_{a,A}\rangle=\langle X_{a,B}\rangle$ and $P(X_{a,A})=P(X_{a,B})$ for $a=1,2$. Notice that this does not imply a distribution $P(X_1)$ that is symmetric around $X_1=0$: its remaining asymmetry, and hence the finite uniform displacement $\langle X_{1,A}\rangle=\langle X_{1,B}\rangle$, results from the linear coupling of $X_1$ to $n_1-1$.

Finally, in the high-symmetry metallic phase ($\beta=0.3$), the stochastic trajectories become centered around the undistorted configuration, exploring a broad region around $X_1^{\rm stag}=X_2^{\rm unif}=0$. Consistently, the distributions $P(X_{2,A})$ and $P(X_{2,B})$ collapse into broad distributions centered around zero, signaling the restoration of the high-symmetry state. This behavior contrasts directly with the coherent lattice free-energy landscape, whose minima remain at finite tilting distortion at the same temperature. The stochastic dynamics therefore restores the undistorted phase even in a regime where the coherent lattice approximation still predicts a low-symmetry structure.

We note that, while the comparison between the coherent-lattice free-energy landscape and the stochastic trajectories is illustrative, it should not be overinterpreted. The $Z_2\times Z_2$-symmetric coherent-lattice free-energy landscape describes the free energy of the global order parameters, while an individual stochastic trajectory experiences a local effective potential determined by its environment, through the neighboring lattice expectation values and the self-consistent DMFT bath. In a symmetry-broken phase, this local potential is biased and generally has a single dominant minimum; the persistence of symmetry breaking is therefore not related to an insufficient simulation time for hopping between equivalent minima of the global free-energy landscape. Additional metastable local minima, corresponding to structural defects, may nevertheless occur, as discussed for the single-mode Holstein model in Ref.~\cite{Picano2023_CDW}. While no indication of an appreciably occupied secondary minimum is found in the present equilibrium simulations, such metastable configurations may become important after photoexcitation and provide a mechanism for long-lived structural disorder. A detailed analysis of this possibility is left for future work.

\subsection{Electronic structure across the two-step transition}
\label{sec:results_spectra}

The distinct structural roles of the dimerization and tilting distortions are directly reflected in the electronic spectrum. Figure~\ref{fig:Semiclassical} shows the ensemble-averaged spectral functions
\begin{equation}
    \langle A(\omega)\rangle
    =
    -\frac{1}{\pi}
    \left\langle
    \mathrm{Im}\,
    G^R(\omega)
    \right\rangle_{\rm traj},
\end{equation}
averaged over the $64$ stochastic trajectories, together with the corresponding occupied spectral weight $\langle A(\omega)F(\omega)\rangle$ for representative temperatures in the three equilibrium phases.

In the I phase, the combined effect of the two structural distortions produces an insulating electronic structure. The staggered dimerization splits band $1$ into strongly inequivalent sublattice components, opening a gap at the chemical potential, while the finite tilting distortion enhances the orbital splitting and shifts band $2$ to higher energies. As a result, the spectral weight at the Fermi level is strongly suppressed and the system is insulating.

The intermediate low-symmetry metallic phase clearly illustrates the distinct electronic roles of the two lattice modes. Once the dimerization has melted, the two sublattices of band $1$ become equivalent and form a broad spectral feature crossing the chemical potential, restoring metallic behavior. At the same time, the finite tilting distortion continues to generate a substantial orbital splitting, so that band $2$ remains predominantly unoccupied. The resulting state is therefore metallic while retaining the characteristic electronic signature associated with the tilting distortion.

At still higher temperature, the disappearance of the tilting distortion in the stochastic solution removes the additional crystal-field splitting. The spectra of the two bands increasingly overlap and thermal broadening becomes substantial, producing the broad metallic spectrum characteristic of the high-symmetry metallic phase.

The evolution of the spectra provides a microscopic interpretation of the two-step structural transition. The collapse of the dimerization restores spectral weight at the Fermi level, driving the transition from the insulating phase to the intermediate low-symmetry metallic state, while the persistence of the tilting distortion maintains a sizable orbital splitting between the two bands. Only when the tilting distortion finally disappears do the spectra of the two orbitals become nearly equivalent, signaling the restoration of the high-symmetry phase.

Together with the equilibrium phase diagram, these results show that the restricted Hartree mean-field description already captures the emergence of the intermediate low-symmetry metallic phase. For the present choice of parameters, however, the restoration of the high-temperature phase is obtained only within the full stochastic semiclassical treatment. The present approach therefore provides a unified microscopic description of the interplay between correlated electrons and fluctuating lattice degrees of freedom across the insulating, low-symmetry metallic, and high-symmetry metallic regimes.

\begin{figure*}
    \centering
    \includegraphics[width=1\linewidth]{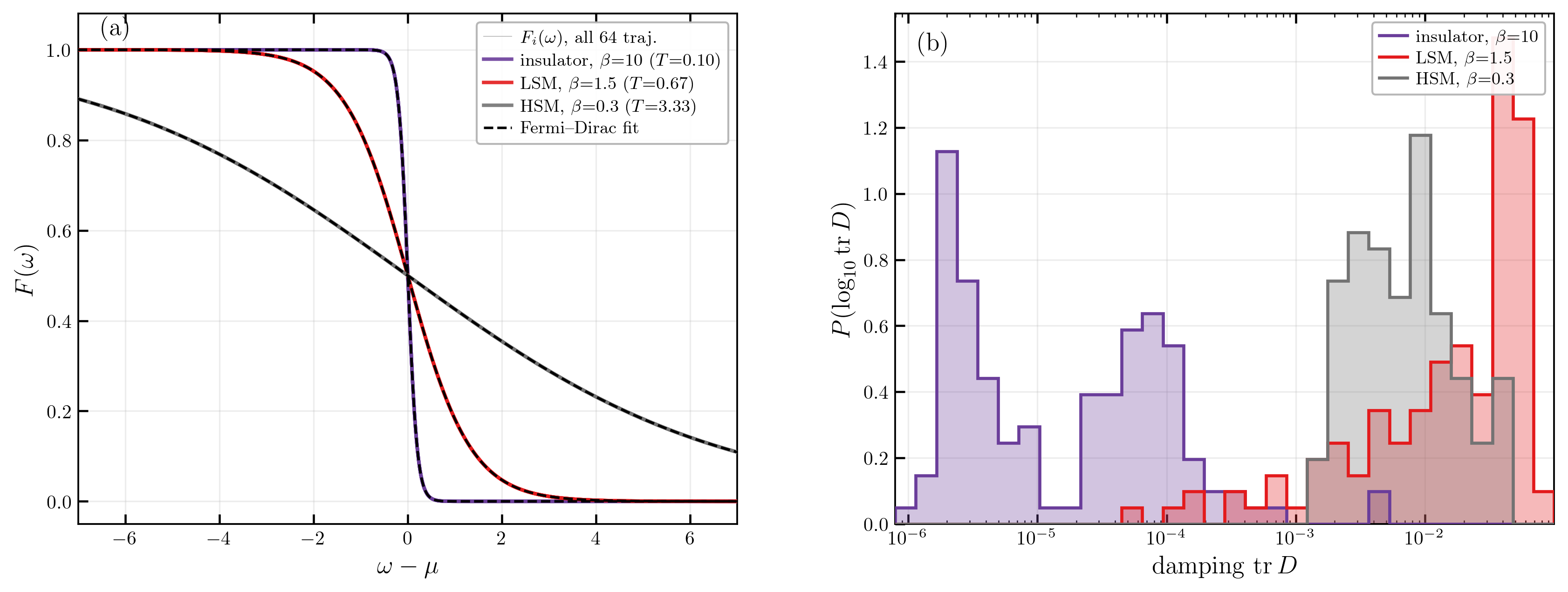}
    \caption{\textbf{Thermal electronic distribution and fluctuating electronic damping.} \textbf{(a)} Local electronic distribution function $F(\omega)$ of all $64$ trajectories in the three equilibrium phases taken at the last time step of the equilibration dynamics. Thin grey lines show the individual distributions $F_i(\omega)$, coloured lines the ensemble average in each phase, and the dashed black line the Fermi--Dirac distribution at the bath temperature. Fitting the slope at the Fermi level, $\beta_i=-4\,\partial_\omega F_i(\omega)|_{\omega=0}$, returns the bath value with a trajectory-to-trajectory spread below $10^{-4}$ in all three phases. The distribution is propagated by the QBE rather than constrained to remain thermal, so the agreement with the Fermi--Dirac form is a result rather than an input. \textbf{(b)} Distribution of $\log_{10}\mathrm{tr}\,\mathbf D$ over the stochastic ensemble for the same three phases taken at the last time step of the equilibration dynamics. In contrast to the electronic distribution function, the damping varies by orders of magnitude between trajectories, reflecting the configuration-dependent electronic back-action on the lattice.}
    \label{fig:supp_FDT}
\end{figure*}

\subsection{Role of electronic fluctuations in the stochastic semiclassical approach}\label{sec:results_born_oppenheimer}

The SCA goes beyond an adiabatic Born--Oppenheimer description in two distinct ways: (i) The electronic distribution function $F_{j,a}(\omega,t)$ for each trajectory is evolved dynamically through the QBE and is therefore in principle not constrained to remain thermal. (ii) The electronic back-action on the lattice is not restricted to the instantaneous Born--Oppenheimer force, but includes the electronic damping and stochastic force of Eqs.~\eqref{eq:sca_damping} and \eqref{eq:sca_noise}, which depend separately for each trajectory on the instantaneous electronic and lattice configuration. The first point turns out to be not relevant for the present equilibrium calculations: As long as the distribution function in the QBE \eqref{scatt_int} remains diagonal in the orbital index, the Fermi--Dirac distribution remains a fixed point of the dynamics even for temporally fluctuating on-site energies: for a thermal distribution, the corresponding self-energies satisfy the equilibrium fluctuation--dissipation relations and the scattering integral vanishes. Figure~\ref{fig:supp_FDT}(a) illustrates this property for the stochastic ensemble in the three phases discussed above. All trajectories remain at the Fermi--Dirac distribution of the bath, with negligible trajectory-to-trajectory variation. A dynamical treatment of the electronic distribution becomes relevant for explicitly nonequilibrium electronic excitations, such as photoexcitation, which we leave for future work.

The second point, by contrast, remains important even in thermal equilibrium. Figure~\ref{fig:supp_FDT}(b) shows the distribution of $\mathrm{Tr}\,\mathbf D$ [Eq.~\eqref{eq:sca_damping}] over the same stochastic ensemble. In all three phases, the damping exhibits pronounced trajectory-to-trajectory fluctuations, spanning several orders of magnitude. A similar behavior is obtained for the noise amplitude, Eq.~\eqref{eq:sca_noise} (not shown here). Thus, although the electronic distribution is thermal and essentially trajectory independent,  the fluctuating electronic feedback on the local lattice coordinates $X_j$ remains strongly configuration dependent. This suggests that retaining trajectory-dependent damping and noise is an important ingredient of the SCA, rather than replacing them by a single phenomenological damping and noise strength.

\section{Conclusions}
\label{sec:conclusions}

We have developed a stochastic semiclassical extension of dynamical mean-field theory for correlated multiorbital systems with nonlinear electron--phonon interactions. Building on the stochastic formalism introduced in Refs.~\cite{Picano2023_CDW,Picano2023_PRB}, we generalized the theory from a single linearly coupled phonon mode to nonlinear electron--phonon couplings involving multiple lattice coordinates. The resulting framework determines self-consistently the deterministic forces, electronic damping, and correlated stochastic noise acting on each lattice coordinate directly from the interacting electronic subsystem, thereby providing a unified microscopic framework for the coupled dynamics of correlated electrons and fluctuating lattice degrees of freedom.

As a first application, we studied a minimal two-orbital model for VO$_2$ containing the two symmetry-distinct structural distortions of the low-temperature insulating phase. The tilting distortion primarily controls the orbital polarization through the crystal-field splitting, while the dimerization opens the insulating gap within the low-energy band. While the bare lattice potential would at most give rise to a transition in which both lattice coordinates simultaneously undergo a displacement, the stochastic semiclassical theory reproduces  three distinct equilibrium phases: the insulating phase, an intermediate low-symmetry metallic phase, and the high-symmetry metal. The lower-temperature transition from the insulator to the low-symmetry metal is primarily electronically driven. It is qualitatively already captured by a more conventional coherent lattice approximation, although the lattice fluctuations kept in the SCA quantitatively renormalize the transition temperature. The higher-temperature restoration of the symmetric phase, in contrast, requires lattice fluctuations and their entropy. The resulting sequence of phases demonstrates how electronic correlations and lattice entropy can act on different structural degrees of freedom and thereby qualitatively reshape the phase diagram of a coupled electron--lattice system.

The present work establishes a stochastic semiclassical framework for correlated electron--lattice systems with nonlinear couplings and multiple structural modes. Since the electronic subsystem is treated within nonequilibrium DMFT using the quantum Boltzmann equation, the formalism is directly applicable to driven and photoexcited systems with nonthermal electronic distributions. A natural next application is therefore the microscopic simulation of the ultrafast photoinduced insulator--metal transition in VO$_2$, where experiments have revealed rapid lattice disordering and transient inhomogeneously disordered states~\cite{Wall2018,Johnson2023_NatPhys}. The present framework is ideally suited to investigate these nonequilibrium phenomena by treating correlated electronic dynamics and stochastic lattice fluctuations self-consistently throughout the photoinduced evolution~\cite{Yang2020}.

\section{Acknowledgments}
We are thankful to S. Wall for a critical reading of our manuscript and for insightful comments on it. We acknowledge stimulating discussions with M. Schir\`o, D.M. Kennes, R. Thomale, A.S. Johnson, A. Zong, A. Kogar, M. Trigo, and J. Del Valle.  A. P. acknowledges funding from the European Union’s Horizon 2020 research and innovation programme under the Marie Skłodowska-Curie Postdoctoral Fellowship (Grant Agreement No. 101149691, DISRUPT). M.E.  acknowledges funding through the Deutsche Forschungsgemeinschaft through OPTIMAL-FOR5750 - 531215165 (Project P1),  QUAST- FOR5249-449872909 (Project P6), and through the Cluster of Excellence ``CUI: Advanced Imaging of Matter'' of the Deutsche Forschungsgemeinschaft (DFG) – EXC 2056, project ID 390715994. A.P. acknowledges the use of computational resources provided by the Collège de France IPH cluster and the Adastra supercomputer hosted at CINES. A.P. also thanks Philipp Hansmann and the RRZE at Friedrich-Alexander University Erlangen-Nürnberg for generously providing additional computational resources. 

\appendix

\section{Derivation of the stochastic semiclassical equations} \label{app:derivation}

In this Appendix, we derive the stochastic semiclassical equations summarized in Sec.~\ref{sec:sca_summary}. We first formulate the coupled electron--lattice problem on the Keldysh contour and integrate out the electronic degrees of freedom to obtain an effective action for the lattice fields. We then expand the resulting action in the quantum components of the lattice fields and perform the semiclassical truncation. Taking the white-noise limit appropriate when the electronic correlation time is short compared with the lattice dynamics yields the stochastic equations of motion. Finally, we specialize these equations to the two-mode VO$_2$ model introduced in Sec.~\ref{sec:model}.

\subsection{Action-based formalism}
\label{sec:action}
We aim to solve the problem using a path integral formulation on the Keldysh contour $\mathcal{C} = \mathcal{C}^+ \cup \mathcal{C}^-$, with $\mathcal{C}^+ = ( - t_{\rm max},  t_{\rm max})$ and $\mathcal{C}^- = (  t_{\rm max}, - t_{\rm max})$, $t_{\rm max}\to\infty$.  We will henceforth use a notation such that the contour time $\tau=t^{+}$ ($\tau=t^{-}$) denotes the physical time $t$ on the upper (lower) branch of $\mathcal{C}$. In order to derive stochastic equations of motion for the phonon variables,  we keep the discrete-time notation for the path integral, and divide $\mathcal{C}$ into $2N-2$ time intervals of length $\delta_t$. For this derivation, the site indices are not relevant, thus we will omit them in this section (we reintroduce them when considering the continuous time limit). Similarly to \cite{Picano2023_PRB}, both the phonons $X_1$ and $X_2$ are local. In the following, we use the notation $X_a$, with $a=1,2$. The physical time $t$ takes $N$ equidistant values $t_1=-t_{\rm max}, ... , t_N=t_{\rm max}$, while the contour times $\tau_m$ run from $m=1,...,2N$, with  $\tau_1=-t_{\rm max}^+$, $\tau_N=t_{\rm max}^+$, $\tau_{N+1}=t_{\rm max}^-$, and $\tau_{2N}=-t_{\rm max}^-$. The action is given in terms of the displacement field $X_{a, m}=X_{a} (\tau_m)$, and Grassmann fields $c_{a, m}=c_a (\tau_m)$, $\bar c_{a, m}=\bar c_a (\tau_m)$ for the electrons, $a = 1,2$. In addition to the time index $m$, electron operators carry spin indices, which are not shown for simplicity. It is convenient to represent the fields $X_{a, m}$ in terms of their so-called quantum and classical components, which are functions of the physical time $t\in(-\infty,\infty)$. We first denote by $X^{\pm}$ the fields on the upper/lower branch as a function of the physical time $t_m$ $(m=1,...,N)$,
\begin{align}
	\label{Xpm}
	&X^{+}_{a, m} \equiv X_{a} (t_m^+) = X_{a, m},  
	\\
	\label{Xpm2}
	&X^{-}_{a, m} \equiv X_{a} (t_m^-) = X_{a, 2N+1-m} . 
\end{align}
Quantum and classical components are introduced by the transformation,
\begin{align}
	\begin{pmatrix}
		X^{\text{cl}}_{a, m} \\
		X^{\text{q}}_{a, m}
	\end{pmatrix} = 
	\frac{1}{2} \begin{pmatrix}
		X^+_{a, m} + X^-_{a, m} \\
		X^+_{a, m} - X^-_{a, m}
	\end{pmatrix} .
\end{align}
By integrating out the electrons, one gets an effective action for the phonons:
\begin{align}
	&e^{i S_{\rm eff}[X_1, X_2 ]}
	=
	e^{i S_{x}[X_1, X_2] + i\Gamma[X_1, X_2]},\,\,\,
	\\
	&\Gamma[X_1, X_2]=-i\log\Big\langle
	e^{iS_{cx}[\bar c,c,X_1, X_2]}
	\Big\rangle_{cc},
\end{align}
with $S_{x}[X_1, X_2]$ the purely phononic action related to the Hamiltonian Eq.~(\ref{eq:ham_ph}), while 
\begin{align}
	& S_{cx}= + \sqrt{2 \Omega} g \delta_t \sum_m \Big[ (O_{1,m+1}^+ - O_{1,m-1}^-) X_{1,m}^{\text{cl}} \nonumber \\
	& + (O_{1,m+1}^+ + O_{1,m-1}^-) X_{1,m}^{\text{q}} \Big] \nonumber \\
	& + \frac{\Omega \Delta}{2} \delta_t \sum_m \Big[ (O_{2,m+1}^+ - O_{2,m-1}^-) ( X_{2,m}^{\text{cl},2} + X_{2,m}^{\text{q},2} ) \nonumber \\
	& + 2 (O_{2,m+1}^+ + O_{2,m-1}^-) X_{2,m}^{\text{cl}} X_{2,m}^{\text{q}} \Big] + b.t. ,
\end{align}
is the action for the electron-phonon interaction. The overall signs follow from $S = \int_\mathcal{C} dt \, (-H_{\text{int}})$ with $H_{\text{int}}$ the electron-phonon terms of Eq.~\eqref{eq:ham_loc}. The abbreviation $b.t.$ denotes terms involving only the fields on the boundary of the contour, which will not be important in the following. We use the notation $O^+_{a,m}\equiv O_a (t_m^+)=O_{a,m}$, and $O^-_{a,m}\equiv O_a (t_m^-)=O_{a,2N+1-m}$.

The effective potential $\Gamma$ is expanded in a Taylor series in the quantum variable
\begin{align}
	& \Gamma [X_{1}^{\text{cl}}, X_{1}^{\text{q}}, X_{2}^{\text{cl}}, X_{2}^{\text{q}}] = \sum_{n=0}^\infty \frac{1}{n!} \sum_{a_1, ... , a_n} \sum_{m_1, ... ,m_n} \nonumber \\
    & X^{\text{q}}_{a_1, m_1}\cdots X^{\text{q}}_{a_n, m_n} \tilde \Pi_{a_1, ..., a_n; m_1,...,m_n},
	\label{eq:seriesgamma}
\end{align}
\begin{align}
	& \tilde \Pi_{a_1, ..., a_n; m_1,...,m_n} \\
	& =\frac{-i\,\partial^n }{\partial X^{\text{q}}_{a_1, m_1}\cdots \partial X^{\text{q}}_{a_n, m_n}}\log\Big\langle
	e^{iS_{cx}[\bar c,c,X]}
	\Big\rangle_{cc} \Big|_{X^{\text{q}}=0}.  \nonumber
\end{align}
The interpretation of the coefficients $\tilde \Pi$ is provided by introducing the action:
\begin{align}
    S_\text{cl} & = S_{cc} + \sqrt{2 \Omega} g \delta_t \sum_m (O_{1,m+1}^+ - O_{1,m-1}^-) X_{1,m}^\text{cl} \nonumber \\
    & + \frac{\Omega \Delta}{2} \delta_t \sum_m (O_{2,m+1}^+ - O_{2,m-1}^-) X_{2,m}^{\text{cl},2} + b.t. ,
\end{align}
which describes a purely electronic model subject to the action of the fluctuating fields $X_1^\text{cl} (t)$ and $X_2^\text{cl} (t)$:
\begin{align}
    \hat{H}_{X_1,X_2} (t) = - \sqrt{2 \Omega} g X_1^\text{cl} (t) \hat{O}_1  {\color{blue}-} \frac{\Omega \Delta}{2} X_2^{\text{cl},2} (t) \hat{O}_2 ,
\end{align}
with the time dependent contribution $X_a^\text{cl} (t) = X_{a,m}^\text{cl}$ for $t \in [t_m, t_{m+1}]$. The first order contributions to the expansion \eqref{eq:seriesgamma} read:
\begin{align}
	& \tilde \Pi_{1; m} = + 2 \sqrt{2 \Omega} g \delta_t \langle \bar O_{1, m} \rangle_\text{cl} , \\
	& \tilde \Pi_{2; m} = + 2 \Omega \Delta \delta_t \langle \bar O_{2, m} \rangle_\text{cl} X^{\text{cl}}_{2, m} ,
\end{align}
with $\bar O_{m} = (O^+_{m+1}+O^-_{m-1})/2$. The second order contributions are obtained as:
\begin{align} \label{eq:Pi_second}
	& \tilde \Pi^K_{11; m,l} = - 4 \Omega g^2 \delta_t^2 \big( \chi_{\text{cl}}^K \big)_{11; m,l} , \\
	& \tilde \Pi^K_{12; m,l} = - 2 \sqrt{2} \Omega^{3/2} g \Delta \delta_t^2 \big( \chi_{\text{cl}}^K \big)_{12; m,l} X^{\text{cl}}_{2, l} , \nonumber \\
    &  \tilde \Pi^K_{21; m,l} = - 2 \sqrt{2} \Omega^{3/2} g \Delta \delta_t^2 X^{\text{cl}}_{2, m} \big( \chi_{\text{cl}}^K \big)_{21; m,l} ,\nonumber \\
	& \tilde \Pi^K_{22; m,l} = -\Omega^{2} \Delta^2 \delta_t^2 \Big[ \delta_{m,l} \langle \tilde{O}_{2, m} \rangle_\text{cl}  + 2 X^{\text{cl}}_{2,m} \big( \chi_{\text{cl}}^K \big)_{22; m,l} X^{\text{cl}}_{2,l} \Big] , \nonumber
\end{align}
with
\begin{align} 
\label{eq:keldysh_discr_time}
	\big( \chi_{\text{cl}}^K \big)_{a b; m l} = -2 i \big[ \langle \bar O_{a,m} \bar O_{b,l} \rangle_\text{cl} - \langle \bar O_{a,m} \rangle_\text{cl} \langle \bar O_{b,l} \rangle_\text{cl} \big] ,
\end{align}
and $\tilde{O}_m = (O^+_{m+1}-O^-_{m-1})/2$. $\big( \chi_{\text{cl}}^K \big)_{a b; m l}$ is the discretized Keldysh component:
\begin{align} \label{eq:keldysh_cont_time}
	\big( \chi_{\text{cl}}^K \big)_{a b} (t,t') = - i \big[ \langle \hat O_a (t) \hat O_b (t') \rangle^{\text{con}}_\text{cl} + t \leftrightarrow t'  \big] ,
\end{align}
of the connected autocorrelation function, see Eq.~\eqref{eq:keldysh_discr_time}, of the electrons in the presence of the fluctuating classical field, where the corresponding electronic operators are indicated with the hat. In the continuous limit, $\tilde{O}_m$ goes to zero, and we omit this contribution to Eq.~\eqref{eq:Pi_second} in the following.

Compared with Ref.~\cite{Picano2023_PRB}, the quadratic coupling
produces terms of second order in $X_2^\text{q}$ already in the
electron--phonon action, before any expansion. This is the origin both of
the configuration-dependent coefficients in Eq.~\eqref{eq:Pi_second} and
of the contact contribution $\delta_{m,l} \langle \tilde{O}_{2,m}
\rangle_\text{cl}$, which is absent for a linear coupling and which, as
noted, drops out in the continuum limit.

\subsection{Semiclassical approximation}

The semiclassical approximation corresponds to truncating the expansion \eqref{eq:seriesgamma} at the second order. The first order in the expansion \eqref{eq:seriesgamma} is combined with the free phonon action into 
\begin{align} 
	\label{action11}
	& S_x +\Gamma^{(1)} = -2 \delta_t \sum_{m=2}^{N-1} \boldsymbol{X}_m^{\text{q}} \cdot \Big( \ddot{\boldsymbol{X}}_m^{\text{cl}} \nonumber \\
	&+ \boldsymbol{\nabla}_{X^{\text{cl}}_m} \mathcal{V} (\mathbf{X}_{m}^{\text{cl}}) - \frac{\boldsymbol{\tilde \Pi}_m}{2 \delta_t} \Big) ,
\end{align}
where the bold symbols with a single time index indicate a vector in the phononic components, e.g., $\boldsymbol{\tilde \Pi}_m = (\tilde \Pi_{1; m}, \tilde \Pi_{2; m})$. The quadratic term in Eq.~\eqref{eq:seriesgamma}, $\Gamma^{(2)}[X]=- \frac{\Omega \delta_t^2}{2} \sum_{m,l} \boldsymbol{X}_m^{\text{q}} \cdot \boldsymbol{\tilde \Pi}_{m,l}^K \cdot \boldsymbol{X}^{\text{q}}_l$, having introduced the matrix
\begin{align}
	& \boldsymbol{\tilde \Pi}^K_{m,l} = - \frac{1}{\Omega \delta_t^2} \begin{pmatrix}
		\tilde \Pi^K_{1,1; m,l} & \tilde \Pi^K_{1,2; m,l} 
		\\
		\tilde \Pi^K_{2,1; m,l} & \tilde \Pi^K_{2,2; m,l}
	\end{pmatrix} , \nonumber
\end{align}
is decoupled using a Hubbard-Stratonovich transformation with a real vectorial field $\boldsymbol{\xi}$,
\begin{align}
	\label{hubbard_stratonovich01}
	& e^{i\Gamma^{(2)}}=
	e^{ - i \frac{\Omega \delta_t^2}{2} \sum_{m,l} \boldsymbol{X}^{\text{q}}_m \cdot \ \boldsymbol{\tilde \Pi}^K_{m,l} \cdot  \ \boldsymbol{X}^{\text{q}}_l }
	\nonumber\\
	& = \frac{1}{Z_{\boldsymbol{\xi}}} \int\mathcal{D}[\boldsymbol{\xi}]
	e^{ - \frac{1}{2} \sum_{m,l} \boldsymbol{\xi}_m \cdot  \boldsymbol{A}_{m,l} \cdot \boldsymbol{\xi}_{l} + i 2 \delta_t \sqrt{\Omega}  \sum_{m} \boldsymbol{\xi}_m \cdot \boldsymbol{X}^{\text{q}}_m } ,
\end{align}
with $(\boldsymbol{A}^{-1})_{m,l}=i \boldsymbol{\tilde \Pi}^K_{m,l}/4$. Here $\int\mathcal{D}[\boldsymbol{\xi}] = \int \prod_{m}d \boldsymbol{\xi}_m$, and 
\begin{align}
	Z_{\boldsymbol{\xi}}=\int\mathcal{D}[\boldsymbol{\xi}] e^{-\frac{1}{2} \sum_{m,l} \boldsymbol{\xi}_m \cdot \boldsymbol{A}_{m,l} \cdot  \boldsymbol{\xi}_{l}}
\end{align}
is the normalization factor. By combining all terms which are linear in $\mathbf{X}^{\text{q}}$, we obtain the equation of motion, valid for each time step $m \ge 2$:
\begin{align} \label{eq:eom}
	\ddot{\boldsymbol{X}}_m^{\text{cl}} = - \boldsymbol{\nabla}_{X^{\text{cl}}_m} \mathcal{V} (\mathbf{X}_{m}^{\text{cl}}) + \frac{\boldsymbol{\tilde \Pi}_m}{2 \delta_t} + \sqrt{\Omega} \boldsymbol{\xi}_m .
\end{align}
If the expansion in the quantum variable had been truncated at the leading order, the $\boldsymbol{\xi}$ integral would have been absent, and Eq.~\eqref{eq:eom} would have become the equation of motion obtained from a mean-field decoupling of the electron-phonon interaction. The term $\boldsymbol{\xi}$ can be viewed as a stochastic force, whose statistics is determined by the matrix $A$, which itself depends on the trajectories $X_{1,m}^{\text{cl}}$ and $X_{2,m}^{\text{cl}}$.


\subsection{White noise limit}
The stochastic equations can be further simplified if the electronic timescale $\tau_e$, which is determined by the bandwidth of the bath, is much shorter than the period $1/\Omega$ of the phonon. In this case, we can try to numerically solve for the phonon dynamics on a time grid $\Delta t$ which is sufficiently short compared to $1/\Omega$, but still long compared to $\tau_e$, $\tau_e \ll \Delta t \ll 1/\Omega$. Electronic correlation functions, in particular the autocorrelation function, vanish for time differences $t-t'\gg \tau_e$, which implies that also the noise becomes uncorrelated on these times,
\begin{align}
	\label{llwkfwf}
	\langle
	\xi_a (t_1) \xi_b (t_2)
	\rangle \to 0 \text{~~~for~~} |t_1-t_2|\gg \tau_e .
\end{align}
As a consequence, the phonon dynamics on the coarser time grid $\Delta t$ should be reduced to a stochastic equation with a force that is uncorrelated between different time steps (white noise). In this limit, one arrives at the coupled equations of motion:
\begin{align}
	\mathbf{X}^{\text{cl}}_{m+1} = \mathbf{X}^{\text{cl}}_{m} + \Delta t \mathbf{V}_m , \\
	\mathbf{V}_{m} = \mathbf{V}_{m-1} + \Delta t \mathbf{F}_{m-1} ,
\end{align}
with: 
\begin{align}
	\mathbf{F}_m = & - \boldsymbol{\nabla}_{X^{\text{cl}}_m} \mathcal{V} (\mathbf{X}_{m}^{\text{cl}}) + \frac{\boldsymbol{\tilde \Pi}_m }{2 \delta_t}  - \Omega \boldsymbol{\Gamma}_m \cdot \mathbf{V}_m + \sqrt{\Omega} \boldsymbol{\xi}_m . \nonumber
\end{align}
Here, we have introduced the matrix:
\begin{align}
	\boldsymbol{\Gamma}_m = - \frac{1}{2} \partial_\omega \text{Im} \boldsymbol{\tilde \Pi}^R (t_m, \omega) \vert_{\omega = 0} ,
\end{align}
where $\boldsymbol{\tilde \Pi}^R (t_m, \omega)$ is the Fourier transform of the retarded component of the matrix $\boldsymbol{\tilde \Pi}$ in which the classical distortions are computed at the lower bound of the considered time-interval (in the definitions Eqs.~\eqref{eq:Pi_second}, $\chi^K$ has to be replaced by $\chi^R$), i.e. 
\begin{align}
        \label{FT}
	\boldsymbol{\tilde \Pi}^R(t,\omega) = \int_0^{\infty } ds \,e^{i\omega s} \boldsymbol{\tilde \Pi}^R (t,t-s) .
\end{align} 
In the white noise limit, we have:
\begin{align}
	& \langle \xi_{a,m} \xi_{b, m'} \rangle = - \frac{1}{4 \Delta t} \text{Im} \boldsymbol{\tilde \Pi}^K_{a b} (t_m, \omega=0) \delta_{m,m'} \nonumber ,
\end{align}
Written explicitly, the retarded component of $ \boldsymbol{\tilde \Pi}^R_{m,l}$ reads:
\begin{align}
	& \boldsymbol{\tilde \Pi}^R_{m,l} = - \frac{1}{\Omega \delta_t^2} \begin{pmatrix}
		\tilde \Pi^R_{1,1; m,l} & \tilde \Pi^R_{1,2; m,l} 
		\\
		\tilde \Pi^R_{2,1; m,l} & \tilde \Pi^R_{2,2; m,l}
	\end{pmatrix} , \nonumber
\end{align}
with
\begin{align} \label{eq:Pi_R_second}
	& \tilde \Pi^R_{11; m,l} = - 4 \Omega g^2 \delta_t^2 \big( \chi_{\text{cl}}^R \big)_{11; m,l} , \\
	& \tilde \Pi^R_{12; m,l} = - 2 \sqrt{2} \Omega^{3/2} g \Delta \delta_t^2 \big( \chi_{\text{cl}}^R \big)_{12; m,l} X^\text{cl}_{2, l} , 
 \nonumber \\
 &\tilde \Pi_{21; m,l}^R = - 2 \sqrt{2} \Omega^{3/2} g \Delta \delta_t^2 X^{\text{cl}}_{2, m} \big( \chi_{\text{cl}}^R \big)_{21; m,l}
 \nonumber \\
	& \tilde \Pi^R_{22; m,l} = -\Omega^{2} \Delta^2 \delta_t^2 \Big[ \delta_{m,l} \langle \tilde{O}_{2, m} \rangle_\text{cl}  + 2 X^{\text{cl}}_{2,m} \big( \chi_{\text{cl}}^R \big)_{22; m,l} X^{\text{cl}}_{2,l} \Big] . \nonumber
\end{align}
In the continuous limit, $\tilde{O}_m$ goes to zero, thus we omit this contribution to \eqref{eq:Pi_R_second} in the following.

\subsection{Equations of motion for the VO$_2$ model}
We summarize the equations of motion we have to solve for the specific problem at hand. Introducing back the $j$ index corresponding to the specific impurity, we might write:
\begin{widetext}
\begin{align}
 \label{eq_motion}
	& \ddot{X}^\text{cl}_{1,j} + \Omega^2 X^\text{cl}_{1,j} + 2 \Omega^2 \mu_1 X^{\text{cl},2}_{2,j} X^\text{cl}_{1,j} + \Omega^2 \mu_2 (X^{\text{cl},2}_{1,j} - X^{\text{cl},2}_{2,j}) X^\text{cl}_{1,j} + \Omega^3 \nu (X^{\text{cl},2}_{1,j} + X^{\text{cl},2}_{2,j})^2 X^\text{cl}_{1,j} - f^{\text{MF}}_{1,j} \nonumber \\
	& + \Omega \Gamma_{11,j} \dot{X}^\text{cl}_{1,j} + \Omega \Gamma_{12,j} \dot{X}^\text{cl}_{2,j} + \gamma_{\text{ph}} \dot{X}^\text{cl}_{1,j} - \sqrt{\Omega} \xi_{1,j} - \eta_{1,j} + 2 \Omega J^{\text{ph}} \langle X^\text{cl}_{1,m} \rangle_{\text{nn} j} = 0 , \nonumber \\
	& \ddot{X}^\text{cl}_{2,j} + \Omega^2  X^\text{cl}_{2,j} + 2 \Omega^2 \mu_1 X^{\text{cl},2}_{1,j} X^\text{cl}_{2,j} - \Omega^2 \mu_2 (X^{\text{cl},2}_{1,j} -  X^{\text{cl},2}_{2,j}) X^\text{cl}_{2,j} + \Omega^3 \nu (X^{\text{cl},2}_{1,j} + X^{\text{cl},2}_{2,j})^2 X^\text{cl}_{2,j} - f^{\text{MF}}_{2,j} \nonumber \\
	& + \Omega \Gamma_{22,j} \dot{X}^\text{cl}_{2,j}  + \Omega \Gamma_{21,j} \dot{X}^\text{cl}_{1,j} + \gamma_{\text{ph}} \dot{X}^\text{cl}_{2,j} - \sqrt{\Omega} \xi_{2,j} - \eta_{2,j} - 2 \Omega J^{\text{ph}} \langle X^\text{cl}_{2,m} \rangle_{\text{nn} j} = 0 ,
\end{align}
\end{widetext}
having introduced the mean field forces:
\begin{align}
	\label{eqn:f_MF} 
	& f^{\text{MF}}_{1,j} (t) = + \sqrt{2 \Omega} g \langle n_{1,j} (t) -1 \rangle_\text{cl} , \nonumber \\
	& f^{\text{MF}}_{2,j} (t) = + \Omega \Delta \langle n_{1,j} (t) - n_{2,j}(t) \rangle_\text{cl} X^{\text{cl}}_{2,j} (t) ,
\end{align}

where $ \langle n_{a,j} (t)\rangle_{\text{cl}} =-i G^<_{\text{cl},a,j}(t,t)$ (occupations include the spin sum). The terms $\gamma_{\text{ph}} \dot{X}^\text{cl}_{a,j} - \eta_{a,j}$ describe an optional phenomenological Langevin bath acting directly on the lattice coordinates, which provides the dissipation needed for the lattice to equilibrate; its white noise obeys the classical fluctuation--dissipation relation
\begin{align}
    \langle \eta_{a,j} (t) \eta_{b,j'} (t') \rangle = \frac{2 \gamma_{\text{ph}}}{\beta} \delta_{a b} \delta_{j j'} \delta (t - t') .
\end{align}
The electronic damping contributions are:
\begin{widetext}
\begin{align}
    \label{eqn:Gamma}
	& \Gamma_{11,j} (t) = -2 g^2 \text{Im} \big[ \partial_{\omega} \big( \chi_\text{cl} \big)_{11,j}^R (\omega, t) \big] \vert_{\omega = 0} , \nonumber \\
	& \Gamma_{22,j} (t) = - \Omega \Delta^2 \text{Im} \big[ \partial_\omega \big( X_2^\text{cl} \chi_\text{cl} X_2^\text{cl} \big)_{22,j}^R (\omega, t) \big] \vert_{\omega = 0}  \approx - \Omega \Delta^2 X_2^\text{cl} (t) \text{Im} \big[ \partial_\omega \big(  \chi_\text{cl}  \big)_{22,j}^R (\omega, t) \big] \vert_{\omega = 0} X_2^\text{cl}(t) \nonumber , \\
	& \Gamma_{12,j} (t) = -\sqrt{2 \Omega} g \Delta \text{Im} \big[ \partial_{\omega} \big( \chi_\text{cl} X_2^\text{cl} \big)_{12,j}^R (\omega, t) \big] \vert_{\omega = 0} \approx -\sqrt{2 \Omega} g \Delta \text{Im} \big[ \partial_{\omega} \big( \chi_\text{cl}  \big)_{12,j}^R (\omega, t) \big] \vert_{\omega = 0} X_2^\text{cl}(t) \nonumber , \\
	& \Gamma_{21, j} (t) = - \sqrt{2 \Omega} g \Delta \text{Im} \big[ \partial_{\omega} \big( X_2^\text{cl} \chi_\text{cl} \big)_{21, j}^R (\omega, t) \big] \vert_{\omega = 0} \approx - \sqrt{2 \Omega} g \Delta X_2^\text{cl}(t) \text{Im} \big[ \partial_{\omega} \big(  \chi_\text{cl} \big)_{21, j}^R (\omega, t) \big] \vert_{\omega = 0} 
\end{align}
\end{widetext}
and the statistics of the noises:
\begin{widetext}
    \begin{align}
        \label{eqn:covariance}
	     &\langle \xi_{a,j} \rangle = 0 \nonumber , \\
	     \langle \xi_{1,j} (t) \xi_{1, j'} (t') \rangle &= - g^2 \text{Im} \big[ \big( \chi_\text{cl} \big)_{11,j}^K (\omega, t) \big] \vert_{\omega = 0}  \delta_{j j'} \delta (t - t') , \nonumber \\
	     \langle \xi_{2,j} (t) \xi_{2, j'} (t') \rangle &= - \frac{\Omega \Delta^2}{2} \text{Im} \big[ \big( X_2^\text{cl} \chi_\text{cl} X_2^\text{cl} \big)_{22,j}^K (\omega, t) \big] \vert_{\omega = 0}  \delta_{j j'} \delta (t - t') \nonumber \\ &\approx - \frac{\Omega \Delta^2}{2} X_2^\text{cl}(t) \text{Im} \big[ \big(  \chi_\text{cl}  \big)_{22,j}^K (\omega, t) \big] \vert_{\omega = 0}  X_2^\text{cl}(t) \delta_{j j'} \delta (t - t') \nonumber \\
         \langle \xi_{1,j} (t) \xi_{2, j'} (t') \rangle &= - \sqrt{\frac{\Omega}{2}} g \Delta \text{Im} \big[ \big( \chi_\text{cl} X_2^\text{cl} \big)_{12,j}^K (\omega, t) \big] \vert_{\omega = 0}  \delta_{j j'} \delta (t - t') \nonumber \\
         &\approx- \sqrt{\frac{\Omega}{2}} g \Delta \text{Im} \big[ \big( \chi_\text{cl}  \big)_{12,j}^K (\omega, t) \big] \vert_{\omega = 0} X_2^\text{cl}(t)  \delta_{j j'} \delta (t - t') \nonumber \\
         \langle \xi_{2,j} (t) \xi_{1, j'} (t') \rangle &= - \sqrt{\frac{\Omega}{2}} g \Delta \text{Im} \big[ \big( X_2^\text{cl} \chi_\text{cl} \big)_{21, j}^K (\omega, t) \big] \vert_{\omega = 0} \delta_{j j'} \delta (t - t') \nonumber \\
         &\approx - \sqrt{\frac{\Omega}{2}} g \Delta X_2^\text{cl}(t) \text{Im} \big[ \big(  \chi_\text{cl} \big)_{21, j}^K (\omega, t) \big] \vert_{\omega = 0} \delta_{j j'} \delta (t - t').
    \end{align}
\end{widetext}
The $a,b$ components of $ \big (\chi_\text{cl} \big)_{ab, j}^K$ are defined in Eq.~\eqref{eqn:chi_K_11}-\eqref{eqn:chi_K_22}. The approximations introduced in Eq.~\eqref{eqn:covariance} come from the white noise limit and are explainded in Sec.~\ref{app:white_noise} of the Appendix. The last term in Eq.~\eqref{eq_motion} is an average over the nearest neighbors $m$ of the site $j$, more explicitly written as:
\begin{align}
    \label{eqn:Xavg}
    \gamma_a \langle X^\text{cl}_{a,m} \rangle_{\text{nn} j} 
    &= \frac{\gamma_a}{Z} \sum_{m NN j} X^\text{cl}_{a,m}
\end{align}
with $a = 1,2$, and $\gamma_a=1$ if $a=1$ and $\gamma_a=-1$ if $a=2$. The average in the last line of Eq.~\eqref{eqn:Xavg} has to be taken over all the atoms of the opposite sublattice (same band $a$). The explicit expressions for the retarded and Keldysh susceptibilities in terms of the local Green's functions are given in Appendix~\ref{app:susceptibilities}.

\section{Electronic susceptibilities}
\label{app:susceptibilities}

In this Appendix, we give the explicit bubble expressions for the electronic response functions entering the damping matrix and noise covariance introduced in Eqs.~\eqref{eq:sca_damping} and~\eqref{eq:sca_noise}. Starting from the general susceptibility defined in Appendix~\ref{app:derivation} and using the local operators $O_1=n_1-1$ and $O_2=n_1-n_2$, the retarded components are
\begin{align}
    \label{eqn:chi_R_11}
	\chi_{\text{cl},11,j}^R (t,t') = - i & \big[ G^R_{\text{cl},1,j} (t,t') G^K_{\text{cl},1,j} (t',t) \nonumber \\
    & + G^K_{\text{cl},1,j} (t,t') G^A_{\text{cl},1,j} (t',t) \big] ,
\end{align}
\begin{align}
	\chi_{\text{cl},12,j}^R (t,t') = \chi_{\text{cl},21,j}^R (t,t') = \chi_{\text{cl},11,j}^R (t,t')
\end{align}

\begin{align}
    \label{eqn:chi22}
	& \chi_{\text{cl},22,j}^R (t,t') = \chi_{\text{cl},11,j}^R (t,t') \\
    & - i \big[ G^R_{\text{cl},2,j} (t,t') G^K_{\text{cl},2,j} (t',t) + G^K_{\text{cl},2,j} (t,t') G^A_{\text{cl},2,j} (t',t) \big] . \nonumber
\end{align}
Similarly, for the Keldysh component, we write: 
\begin{align}
    \label{eqn:chi_K_11}
	\chi_{\text{cl},11,j}^K (t,t') = - 2 i & \big[ G^<_{\text{cl},1,j} (t,t') G^>_{\text{cl},1,j} (t',t) \nonumber \\
    & + G^>_{\text{cl},1,j} (t,t') G^<_{\text{cl},1,j} (t',t) \big] , \nonumber \\
\end{align}
\begin{align}
	\chi_{\text{cl},12,j}^K (t,t') = \chi_{\text{cl},21,j}^K (t,t') = \chi_{\text{cl},11,j}^K (t,t')
\end{align}
\begin{align}
        \label{eqn:chi_K_22}
	& \chi_{\text{cl},22,j}^K (t,t') = \chi_{\text{cl},11,j}^K (t,t')\\
    & {- 2 i} \big[ G^<_{\text{cl},2,j} (t,t') G^>_{\text{cl},2,j} (t',t) + G^>_{\text{cl},2,j} (t,t') G^<_{\text{cl},2,j} (t',t) \big] \nonumber .
\end{align}

We notice that the mixed orbitals Green's functions are zero in this setup. All Green's functions above are \emph{per spin}, while the operators $O_a$ contain the spin summation: every susceptibility bubble therefore carries an overall factor $2$. In the retarded expressions, Eqs.~\eqref{eqn:chi_R_11}--\eqref{eqn:chi22}, this factor is generated automatically by the Keldysh-function form of the bubble; in the Keldysh components above we write the factor $2$ explicitly.

The Green's function matrix for the $a$ orbitals reads \footnote{We should have: $\Sigma^{(2)}_{a,\sigma} (t,t') = U(t) U(t') G_{a,\sigma} (t, t') \big( G_{a,\bar{\sigma}} (t', t) G_{a,\bar{\sigma}} (t, t') + \sum_{\sigma'} G_{\bar{a},\sigma'} (t',t) G_{\bar{a},\sigma'} (t,t') \big)$.}:
\begin{align} \label{Gloc}
	G^{-1}_{\text{cl},a,j} (t,t') & = [ (i \partial_t + \mu) - h_{a,j}(t)] \delta_{\mathcal{C}} (t,t') \nonumber \\ &  - \Delta_{a,j} (t,t') -\Sigma_{a,j} (t,t') ,
\end{align}
with 
\begin{align} 
\label{hloc}
	h_{1,j}(t) & = e_{1,j} -g \sqrt{2 \Omega}X_{1,j} - \Omega \frac{\Delta}{2} X^2_{2,j} , \nonumber \\
	h_{2,j}(t) & = \Omega \frac{\Delta}{2} X^2_{2,j}  .
\end{align}
$e_{1,j}$ is positive if $j \in A$ and negative if $j \in B$. It is a staggered field which is turned on only for the first DMFT iteration at equilibrium in order to allow the $A$ and $B$ to eventually break the symmetry.

\section{Coherent lattice approximation}
\label{app:meanfield}
In this Appendix, we derive the coherent-lattice mean-field equations and construct the reference equilibrium free-energy landscape used in Sec.~\ref{sec:results}. Starting from the model introduced in Sec.~\ref{sec:model}, we neglect lattice fluctuations and restrict the distortion pattern to a staggered dimerization and a uniform tilt. Setting the damping and the noise in Eq.~\eqref{eq_motion} to zero --- that is, truncating the expansion Eq.~\eqref{eq:seriesgamma} at first order in the quantum component --- reduces the lattice dynamics to the Ehrenfest form $\ddot{\mathbf{X}} = - \boldsymbol{\nabla} \mathcal{V} + \mathbf{f}^\text{MF}$, whose stationary solutions $\boldsymbol{\nabla} \mathcal{V} = \mathbf{f}^\text{MF}$ define the self-consistent Born--Oppenheimer configurations. In this section we solve these stationarity conditions, with two further simplifications which allow us to obtain them from an explicit free-energy functional: the Hubbard interaction is decoupled at the Hartree level, and the distortion pattern is restricted to $X_{1,B} = - X_{1,A}$ and $X_{2,B} = X_{2,A}$, i.e. to a staggered dimerization and a uniform tilt. The resulting $F_\text{MF}$, Eq.~\eqref{eq:F_MF}, provides the landscape against which the stochastic trajectories of Sec.~\ref{sec:results} are compared; no analogous functional exists for the full stochastic problem, in or out of equilibrium. As we show in Sec.~\ref{sec:results_phasediagram}, releasing the restriction on the distortion pattern is quantitatively important already in equilibrium.
 
The solution of the problem is achieved by solving self-consistently the phononic and the electronic problem, where the Hubbard interaction is decoupled with the mean-field ansatz:
\begin{align}
    & \langle n_{A,1} \rangle = \frac{\bar{n}}{2} + \frac{\delta n}{2} + \frac{m_\text{o}}{4} , \nonumber \\
    & \langle n_{A,2} \rangle = \frac{\bar{n}}{2} - \frac{m_\text{o}}{4} , \nonumber \\
    & \langle n_{B,1} \rangle = \frac{\bar{n}}{2} - \frac{\delta n}{2} + \frac{m_\text{o}}{4} , \nonumber \\
    & \langle n_{B,2} \rangle = \frac{\bar{n}}{2} - \frac{m_\text{o}}{4} ,
\end{align}
where $A$ and $B$ ($1$ and $2$) are sublattice (orbital) indexes, $\bar{n} = 1$ is the average occupation per site, $\delta n$ is the charge-density wave order parameter related to the activation of the Holstein phonon $X_1$ ($\delta n = \langle n_{A,1} \rangle - \langle n_{B,1} \rangle$) and $m_\text{o}$ is the orbital imbalance order parameter related to the activation of the phonon $X_2$ ($m_\text{o} = \langle n_{A,1} \rangle + \langle n_{B,1} \rangle - \langle n_{A,2} \rangle - \langle n_{B,2} \rangle$). The mean-field decoupled Hubbard interaction becomes:
\begin{align}
    & \frac{U}{2} \big[ n_A (n_A - 1) + n_B (n_B - 1 ) \big] \nonumber \\
    &  \approx \frac{U}{4} \Big[ 3 \bar{n} \ (n_A + n_B) - 3 \bar{n}^2 - \frac{1}{4} \delta n^2 + \frac{1}{4} m_\text{o}^2  \nonumber \\
    & + \delta n \ (n_{A,1} + 2 n_{A,2} - n_{B,1} - 2 n_{B,2}) \nonumber \\
    & - \frac{m_\text{o}}{2} (n_{A,1} + n_{B,1} - n_{A,2} - n_{B,2}) \Big] .
\end{align}
By moving to the energy representation of the problem and by neglecting the interband hopping (its contribution is crucial only out-of-equilibrium to allow a transfer of charge from one band to the other), we get the mean-field Hamiltonian:
\begin{align} \label{eq:mf_ham}
    H_\textbf{MF} & = \sum_{a,\sigma} \int d \epsilon \psi^\dagger_{a, \epsilon,\sigma} \begin{pmatrix}
		h_{A,a} & \epsilon 
		\\
		\epsilon & h_{B,a}
	\end{pmatrix} \psi_{a, \epsilon,\sigma} \mathcal{D}_a (\epsilon) \nonumber \\
    & - \frac{3}{4} U \bar{n}^2 - \frac{U}{16} \delta n^2 + \frac{U}{16} m_\text{o}^2 + 2 \mathcal{V} (X_1, X_2) \nonumber \\ 
    &-2 \Omega J^{ph} (X_1^2+X_2^2) ,,
\end{align}
where $\psi^\dagger_{a, \epsilon,\sigma} = (c^\dagger_{A,a,\epsilon, \sigma} , c^\dagger_{B,a,\epsilon, \sigma})$ is a spinor in the sublattice index, $\mathcal{D}_a (\epsilon)$ is the DOS defined in Eq.~\eqref{eq:bethe_dos} and the diagonal components of the matrix are:
\begin{align}
    & h_{A,1} = \frac{3}{4} U \bar{n} + \frac{U}{4} \delta n - \frac{U}{8} m_\text{o} - \sqrt{2 \Omega} g X_1 - \frac{\Omega \Delta}{2} X_2^2 - \mu , \nonumber \\
    & h_{B,1} = \frac{3}{4} U \bar{n} - \frac{U}{4} \delta n - \frac{U}{8} m_\text{o} + \sqrt{2 \Omega} g X_1 - \frac{\Omega \Delta}{2} X_2^2 - \mu , \nonumber \\
    & h_{A,2} = \frac{3}{4} U \bar{n} + \frac{U}{2} \delta n + \frac{U}{8} m_\text{o} + \frac{\Omega \Delta}{2} X_2^2 - \mu , \nonumber \\
    & h_{B,2} = \frac{3}{4} U \bar{n} - \frac{U}{2} \delta n + \frac{U}{8} m_\text{o} + \frac{\Omega \Delta}{2} X_2^2 - \mu .
\end{align}
In the derivation of Eq.~\eqref{eq:mf_ham}, we have assumed the $X_1$ variable to take opposite values on the two sublattices, while $X_2$ is assumed to have the same value on both sublattices.

By introducing the three Pauli matrices $\sigma_x$, $\sigma_y$ and $\sigma_z$ together with the $2 \times 2$ identity matrix $\mathbb{1}$, the Hamiltonian Eq.~\eqref{eq:mf_ham} can be easily diagonalized:
\begin{align}
    & \begin{pmatrix}
		h_{A,a} & \epsilon 
		\\
		\epsilon & h_{B,a}
	\end{pmatrix} = h_{A,a} \frac{\mathbb{1} + \sigma_z}{2} + h_{B,a} \frac{\mathbb{1} - \sigma_z}{2} + \epsilon \sigma_x \nonumber \\
    & = U^\dagger_a (\epsilon) \big[ \frac{h_{A,a} + h_{B,a}}{2} \mathbb{1} + E_a (\epsilon) \sigma_z \big] U_a (\epsilon) ,
\end{align}
with $U_a (\epsilon) = e^{i \theta_a (\epsilon) \sigma_y}$ and having introduced the definitions:
\begin{align}
    & E_a (\epsilon) = \sqrt{\Big( \frac{h_{A,a} - h_{B,a}}{2} \Big)^2 + \epsilon^2} , \nonumber \\
    & \cos \big( 2 \theta_a (\epsilon) \big) = \frac{h_{A,a} - h_{B,a}}{2 E_a (\epsilon)} , \nonumber \\
    & \sin \big( 2 \theta_a (\epsilon) \big) = \frac{\epsilon}{E_a (\epsilon)} . \nonumber
\end{align}
Considering the eigenvalues:
\begin{align}
    E_{a , \pm} (\epsilon) = \frac{h_{A,a} + h_{B,a}}{2} \pm E_a (\epsilon) , \nonumber
\end{align}
one arrives writing the mean-field \emph{free} energy that needs to be minimized at fixed particle number as:
\begin{align}
    \label{eq:F_MF}
    F_\text{MF} & = - \frac{1}{\beta} \sum_{\gamma = \pm} \sum_{a,\sigma} \int d \epsilon \, \ln \Big( 1 + e^{- \beta E_{a , \gamma} (\epsilon)} \Big) \mathcal{D}_a (\epsilon) \nonumber \\
    & + 2 \mu \bar{n} - \frac{3}{4} U \bar{n}^2 - \frac{U}{16} \delta n^2 + \frac{U}{16} m_\text{o}^2 + 2 \, \mathcal{V} (X_1, X_2) \nonumber \\ 
    &-2 \Omega J^{ph} (X_1^2+X_2^2) ,
\end{align}
where the integration domain for $\epsilon$ goes from $- W_a/2$ to $W_a/2$. 

We note that $\partial F_\text{MF} / \partial X_a = 0$ reproduces exactly Eq.~\eqref{eq_motion} with $\Gamma_{ab} = \xi_a = \gamma_\text{ph} = \eta_a = 0$ and all time derivatives set to zero, with the mean-field forces Eq.~\eqref{eqn:f_MF} evaluated in the Hartree approximation. The mean-field phase diagram of this section is therefore the stationary limit of the stochastic theory of Sec.~\ref{sec:semiclassical}, within the restricted ansatz and with the Hartree treatment of $U$.

One might explicitly write the self-consistent equations for the CDW order parameter $\delta n$ as:
\begin{align}
    & \delta n = - \frac{1}{2} (U \delta n - 4 \sqrt{2 \Omega} g X_1) \nonumber \\
    & \int d \epsilon \frac{\mathcal{D}_1 (\epsilon)}{E_1 (\epsilon)} \frac{\sinh \big( \beta E_1 (\epsilon) \big)}{\cosh \Big( \beta \frac{h_{A,1} + h_{B,1}}{2} \Big) + \cosh \big( \beta E_1 (\epsilon) \big)} ,
\end{align}
and for the orbital imbalance $m_\text{o}$:
\begin{align} \label{eq:self_orb_imb}
    & m_\text{o} = 2 \int d \epsilon \mathcal{D}_1 (\epsilon) \frac{\cosh \big( \beta E_1 (\epsilon) \big) + e^{- \beta \frac{h_{A,1} + h_{B,1}}{2}}}{\cosh \Big( \beta \frac{h_{A,1} + h_{B,1}}{2} \Big) + \cosh \big( \beta E_1 (\epsilon) \big)} \nonumber \\
    & - 2 \int d \epsilon \mathcal{D}_2 (\epsilon) \frac{\cosh \big( \beta E_2 (\epsilon) \big) + e^{- \beta \frac{h_{A,2} + h_{B,2}}{2}}}{\cosh \Big( \beta \frac{h_{A,2} + h_{B,2}}{2} \Big) + \cosh \big( \beta E_2 (\epsilon) \big)} .
\end{align}
These two equations have to be supplied with the self-consistent relation regarding the conservation of the total charge of the system, i.e., 
\begin{align}
    & \bar{n} = \int d \epsilon \mathcal{D}_1 (\epsilon) \frac{\cosh \big( \beta E_1 (\epsilon) \big) + e^{- \beta \frac{h_{A,1} + h_{B,1}}{2}}}{\cosh \Big( \beta \frac{h_{A,1} + h_{B,1}}{2} \Big) + \cosh \big( \beta E_1 (\epsilon) \big)} \nonumber \\
    & + \int d \epsilon \mathcal{D}_2 (\epsilon) \frac{\cosh \big( \beta E_2 (\epsilon) \big) + e^{- \beta \frac{h_{A,2} + h_{B,2}}{2}}}{\cosh \Big( \beta \frac{h_{A,2} + h_{B,2}}{2} \Big) + \cosh \big( \beta E_2 (\epsilon) \big)} ,
\end{align}
which allows to determine the chemical potential $\mu$. The equations above can be solved for each value of $X_1$ and $X_2$ and for every inverse temperature $\beta$. We notice that, for $U = X_1 = X_2 = 0$, $\delta n = 0$. If we further assume $\beta \rightarrow + \infty$, we get:
\begin{align}
    m_\text{o} = 2 \Big[ \int_{-W_1/2}^\mu d \epsilon \mathcal{D}_1 (\epsilon) - \int_{-W_2/2}^\mu d \epsilon \mathcal{D}_2 (\epsilon) \Big] ,
\end{align}
which is the expected zero-temperature result, i.e., the orbital imbalance, in this case, is only determined by the different bandwidth of the two bands. If we take, instead, the limit $\beta \rightarrow 0^+$, one gets $m_\text{o} \rightarrow 0$, i.e., at infinite temperature, the different bandwidth of the bands becomes irrelevant and no orbital polarization is observed.

\section{Implementation details}
\label{app:implementation}

This Appendix collects technical details of the electronic and stochastic calculations summarized in Secs.~\ref{sec:DMFT_loop} and~\ref{sec:timestepping}. We first give the IPT self-energy and the self-energy associated with the bosonic thermal bath. We then summarize the evaluation of the electronic polarizability and finally discuss the white-noise approximation and the numerical sampling of the correlated stochastic forces.

\subsection{IPT self-energy}
\label{app:ipt}
The iterated perturbation theory (IPT)~\cite{Aoki2014} approximation is  based on a second-order diagrammatic evaluation of the self-energy in terms of the Weiss-field $\mathcal{G}$, that is the free impurity Green's function calculated in DMFT, rather than the bare noninteracting electron propagator:
\begin{align}
	\Sigma_U(t,t')
	=
	U(t)U(t')
	\mathcal{G}(t,t')
	\mathcal{G}(t,t')
	\mathcal{G}(t',t)
	\label{IPT}
\end{align}
Within IPT the impurity self-energy is evaluated to second order in the Weiss Green’s function,
\begin{align} \label{sigma_ipt}
	\Sigma^{(2)}_{a,j,\sigma}(t,t') &= U(t) U(t') \mathcal{G}_{a,j,\sigma} (t, t') \big[ \mathcal{G}_{a,j,\bar{\sigma}} (t', t) \mathcal{G}_{a,j,\bar{\sigma}} (t, t') \nonumber \\
	&+ \sum_{\sigma'} \mathcal{G}_{\bar{a},j,\sigma'} (t',t) \mathcal{G}_{\bar{a},j,\sigma'} (t,t') \big],
\end{align}
where $\Sigma^{(2)}_{a,j}= \sum_{\sigma} \Sigma^{(2)}_{a,j,\sigma}$. Written explicitly for the band A,1:
\begin{align}
    \Sigma^{(2)}_{1,A,j,\sigma}(t,t')
    &= U(t)U(t')\,
    \mathcal{G}_{1,A,j,\sigma}(t,t')
    \mathcal{G}_{1,A,j,\bar{\sigma}}(t',t)
    \nonumber\\
    &\qquad\times
    \mathcal{G}_{1,A,j,\bar{\sigma}}(t,t')
    \nonumber\\
    &\quad
    +2U(t)U(t')\,
    \mathcal{G}_{1,A,j,\sigma}(t,t')
    \mathcal{G}_{2,A,j,\sigma'}(t',t)
    \nonumber\\
    &\qquad\times
    \mathcal{G}_{2,A,j,\sigma'}(t,t').
\end{align}
For the time-translationally invariant NESS loop, the Green’s functions depend only on the relative time. There, the two time arguments $(t,t')$ are replaced by the time difference $t-t'$ which in the following will be called simply $t$. The lesser and retarded components of the electron-electron interaction  $\Sigma_{U}(t)=U^2 \mathcal{G}(t)\mathcal{G}(-t)\mathcal{G}(t)$ read:
\begin{align}
    \label{Sigma_U}
    \Sigma_{U}^{<(>)}(t) &=U^2(t)[\mathcal{G}(t)\mathcal{G}(-t)]^{<(>)}\mathcal{G}^{<(>)}(t) \nonumber \\
    &=U^2(t)\mathcal{G}^{<(>)}(t)\mathcal{G}^{>(<)}(-t)\mathcal{G}^{<(>)}(t) \nonumber \\
    &=U^2(t)\mathcal{G}^{<(>)}(t)[-\mathcal{G}^{>(<)}(t)]^\dagger \mathcal{G}^{<(>)}(t) \\
    \Sigma_{U}^R(t) &= \Sigma_{U}^>(t) - \Sigma_{U}^<(t)+ \Sigma_{U}^A(t) \nonumber \\
    &= U^2(t)\mathcal{G}^>(t)[-\mathcal{G}^<(t)]^\dagger \mathcal{G}^>(t)- \Sigma_{U}^<(t), \,\,\,\,\,\,  \forall t>0.
    \label{Sigma_U_ret_ss}
\end{align} 
In practice, in the NESS loop, we start from $\mathcal{G}(\omega)$, anti-Fourier transform it in $\mathcal{G}(t)$, get $\Sigma(t)$ from Eqs.~\eqref{Sigma_U}-\eqref{Sigma_U_ret_ss}, and Fourier transform it to get $\Sigma(\omega)$.

\subsection{Phononic thermal bath}
\label{app:thermal_bath}

The coupling to the thermal bosonic bath is included via a local electron-phonon self-energy $\Sigma^{\text{ph}}$. To model an external heat bath we neglect the electronic back-action on the phonons and retain only the lowest-order electron-phonon self-energy,
\begin{align} \label{sphonon}
	\Sigma^{\text{ph}}_{a,j}(t,t')&=i g_{\text{ph}}^2G_{a,j}(t,t') D_{\text{ph}}(t,t'), 
\end{align} 
where: $G$ is the fully interacting local electron Green's function of site $j$; $g_{\text{ph}}$ measures the electron-phonon coupling strength, and $D_{\text{ph}}$ is the propagator for free bosons with an Ohmic density of states $J(\omega)=\omega/(4\omega_{\text{ph}}^2) \exp(-\omega/\omega_{\text{ph}})$  with exponential cutoff $\omega_{\text{ph}}=0.2$.

The spectral density of the propagator $D_{\text{ph}}$, $A_{\text{ph}}(\omega)=-1/\pi \Im D_{\text{ph}}^R(\omega+i0^+)$, which has spectral weight at positive and negative frequencies, is given by
\begin{align}
	A_{\text{ph}} (\omega)=
	\begin{cases}
		J(\omega), \ \ \ \ \forall \omega>0,	\\
		-J(-\omega), \forall \omega<0.
	\end{cases}	
\end{align}
The occupation function of the bosons is kept in equilibrium with inverse temperature $\beta_i$, $b(\omega)=\frac{1}{e^{\beta_i \omega}-1}$. The temperature of the heat bath is the same as the initial one of the system in equilibrium, $\beta_i$, such that the system will eventually thermalize back to its initial temperature long after the excitation.
The lesser and greater components of $D_{\text{ph}}$ are given by:
\begin{align}
  - \Im{D^<_{\text{ph}}(\omega)}  &= 2 \pi A_{\text{ph}}(\omega) b(\omega) \nonumber \\ 
  - \Im{D^>_{\text{ph}}(\omega)}  &= 2 \pi A_{\text{ph}}(\omega) [1+b(\omega)]
\end{align}

In the following, we suppress the orbital and site indices $a,j$ for
 simplicity. The lesser and greater components of
Eq.~\eqref{sphonon} are
\begin{align}
	\label{Sigma_ph_les}
	\Sigma_{\text{ph}}^{<(>)}(t)&=i g^2_{\text{ph}} G^{<(>)}(t)D_{\text{ph}}^{<(>)}(t), \\
	\Sigma_{\text{ph}}^R(t)&=i g^2_{\text{ph}} [G^<(t)D_{\text{ph}}^R(t)+G^R(t)(D_{\text{ph}}^<(t)+D_{\text{ph}}^R(t))] \nonumber \\
	&=i g^2_{\text{ph}} G^>(t) D_{\text{ph}}^>(t)-\Sigma_{\text{ph}}^<(t), \,\,\,\, \forall t>0  
	\label{Sigma_ph_ret}
\end{align}
where, in the last equality we considered that generally $\Sigma^R(t)=\Sigma^>(t)-\Sigma^<(t)+\Sigma^A(t)$, however, since we restrict the time axis to $t>0$, $\Sigma^A(t)=0$.

\subsection{Electronic polarizability}
\label{app:polarizability}

For completeness, we now give the time-domain expressions used in the
numerical evaluation of the bubble susceptibilities of
Appendix~\ref{app:susceptibilities}. We introduce the single-band
polarizability
\begin{align}
    \Pi(t,t')=i\,G(t,t')G(t',t).
\end{align}
Its Keldysh components are given by
\begin{align}
	\label{Pi_ret}
	\Pi^R(t,t')
	&= i \big [ G^R(t,t')G^<(t',t) +G^<(t,t')G^A(t',t) \big ] \nonumber \\
	&= i \big [ G^R(t,t')G^>(t',t) +G^>(t,t')G^A(t',t) \big ] \nonumber \\
	&= \frac{i}{2} \big [ G^R(t,t')G^K(t',t) +G^K(t,t')G^A(t',t) \big ], \\
	\Pi^<(t,t')
	&= i\,G^<(t,t')G^>(t',t),\\
	\Pi^>(t,t')
	&= i\,G^>(t,t')G^<(t',t),\\
	\Pi^K(t,t')
	&=\Pi^<(t,t')+\Pi^>(t,t').
\end{align}

In the nonequilibrium steady-state (NESS) implementation, the Green's functions are time-translationally invariant, $G(t,t')=G(t-t')$, such that
\begin{align}
    \Pi(t)=i\,G(t)G(-t).
\end{align}
Using the symmetry relations
\begin{align}
    G^{<,>,K}(t,t')^*&=-G^{<,>,K}(t',t),\\
    G^{R}(t,t')^*&=G^{A}(t',t),
\end{align}
the expressions implemented in the code become
\begin{align}
    \label{eqn:PiR}
    \Pi^R(t)
    &=i [G^R(t)G^{<(>)}(-t) + G^{<(>)}(t) G^A(-t)] \nonumber \\
    &= i [G^R(t)[-G^{<(>)}(t)]^\dagger + G^{<(>)}(t) [G^R(t)]^\dagger] \nonumber  \\
    &= \frac{i}{2} [G^R(t)G^{K}(-t) + G^{K}(t) G^A(-t)] \nonumber \\
    &= \frac{i}{2} [G^R(t)[-G^{K}(t)]^\dagger + G^{K}(t) [G^R(t)]^\dagger], \\
    \Pi^{<(>)}(t)
    &= i\,G^{<(>)}(t)G^{>(<)}(-t) \nonumber \\
    &= i\,G^{<(>)}(t)[-G^{>(<)}(t)]^\dagger,\\
    \Pi^K(t)
    &=\Pi^<(t)+\Pi^>(t).
\end{align}

\subsection{The white noise approximation}
\label{app:white_noise}
The electronic quantities of interest for the phononic equations of motions are: $f^{\text{MF}}_a$ in Eq.~\eqref{eqn:f_MF}; the dampings $\Gamma$ in Eq.~\eqref{eqn:Gamma}, and the noise variances $\langle \xi \xi \rangle$ in Eq.~\eqref{eqn:covariance}. The mean field forces are easily calculated by recalling that 
\begin{align}
     \langle n_{a,j} (t)\rangle_{\text{cl}} =-i G^<_{\text{cl},a,j}(t,t)
\end{align}

The electronic autocorrelation function in Eq.~\eqref{eq:keldysh_cont_time}, $(\chi^{K}_{cl})_{ab}(t_1,t_2)$ decays to zero  $\text{ for } |t_1 - t_2| \gg \tau_e$ (with $\tau_e$  the electronic time scale). Thus on this time-scale, also the noise $\xi_a$  is uncorrelated, despite the presence of the $X_2$ factors (if any):
\begin{align}
    \left\langle \xi_a(t_1) \xi_b(t_2) \right\rangle &\to 0 \quad \text{for} \quad |t_1 - t_2| \gg \tau_e.
\end{align}
There are stochastic force contributions:
\begin{align}
    \Delta V_{\xi_a} = \int_{0}^{\Delta t} d t_1 \xi_a(t_1) 
\end{align}
where the noise \(\xi_a(t_1)\) is determined by the matrix $(\chi^{K}_{cl})_{ab}(t_1,t_2) \text{ for } t_1, t_2 \in [0, \Delta t]$ (with factors  $X_2(t_a)$ if $a, b = 2$). To the leading order in \(\Delta t\), we can replace:
\begin{align}
    (\chi^{K}_{cl})_{ab}(t_1,t_2) &\rightarrow (\chi^{K}_{cl,0})_{ab}(t_1,t_2)
\end{align}
obtained from the electronic model with the factors $X^{cl}_{a}(t)$ frozen at $X^{cl}_{a}(0)$  for $t \in [0, \Delta t]$.

In a similar spirit, we approximate:
\begin{align}
    \tilde \Pi^{K}_{22}(t_1, t_2) &= -2 \Omega^2 \Delta^2 \delta_t^2 X^{cl}_2(t_1) (\chi^K_{cl})_{22} (t_1,t_2) X^{cl}_2(t_2) \nonumber \\
    &\approx -2 \Omega^2 \Delta^2 \delta_t^2 X^{cl}_2(0) (\chi^K_{cl,0})_{22} (t_1,t_2) X^{cl}_2(0) \nonumber \\
    &=  (\tilde \Pi^{K}_0)_{22}(t_1, t_2)
\end{align}
and similarly for the other components.

Thus, the noise becomes a statistically independent variable with variance:
\begin{align}
    (\sigma^2_{\Delta V_\xi})_{ab}
    &=
    \Omega \int_0^{\Delta t} dt_1
    \int_0^{\Delta t} dt_2\,
    \langle \xi_a(t_1)\xi_b(t_2)\rangle
    \nonumber\\
    &=
    i\frac{\Omega}{4}
    \int_0^{\Delta t} dt_1
    \int_0^{\Delta t} dt_2\,
    \widetilde{\Pi}^{K}_{ab}(t_1,t_2)
    \nonumber\\
    &=
    i\frac{\Omega}{4}
    \int_0^{\Delta t} dt_1
    \int_{t_1-\Delta t}^{t_1} ds\,
    \widetilde{\Pi}^{K}_{ab}(t_1,t_1-s)
    \nonumber\\
    &\approx
    i\frac{\Omega}{4}
    \int_0^{\Delta t} dt_1
    \int_{t_1-\Delta t}^{t_1} ds\,
    \widetilde{\Pi}^{K}_{0,ab}(t_1,t_1-s)
    \nonumber\\
    &\approx
    i\frac{\Omega}{4}
    \int_0^{\Delta t} dt_1
    \int_{-\infty}^{+\infty} ds\,
    \widetilde{\Pi}^{K}_{0,ab}(t_1,t_1-s)
    \nonumber\\
    &=
    i\frac{\Omega}{4}
    \int_0^{\Delta t} dt_1\,
    \widetilde{\Pi}^{K}_{0,ab}(t_1,\omega=0)
    \nonumber\\
    &\approx
    -\frac{\Omega}{4}\Delta t\,
    \Im \widetilde{\Pi}^{K}_{0,ab}(0,\omega=0).
    \label{eqn:sigma}
\end{align}
The second line follows from the definition of $A^{-1}$ in Eq.~\eqref{hubbard_stratonovich01}; the third line from the substitution $t_2=t_1-s$ and $dt_2=-ds$. In the fourth line we make the approximation $\tilde\Pi^K(t_1,t_1-s)\rightarrow \tilde\Pi^K_{0}(t_1,t_1-s)$, while in the fifth line we make use of the fast decay of the electronic autocorrelation function for $|s| \gg \tau_e$.

\subsubsection{Drawing samples $\xi_1$ and $\xi_2$}

In order to draw the sample $\xi_1$ and $\xi_2$ which appear in the equations of motions for the phonons,  we write the covariance matrix of the noise (for each average time $t$):
\begin{align}
    \hat{\Xi}
    &=
    \begin{bmatrix}
        \langle \xi_{1,j}(t)\xi_{1,j'}(t')\rangle &
        \langle \xi_{1,j}(t)\xi_{2,j'}(t')\rangle \\
        \langle \xi_{2,j}(t)\xi_{1,j'}(t')\rangle &
        \langle \xi_{2,j}(t)\xi_{2,j'}(t')\rangle
    \end{bmatrix},
\end{align}
where 
\begin{align}
    \langle \xi_{1,j}(t)\xi_{1,j'}(t')\rangle
    &=
    -\delta_{jj'}\delta(t-t')\,g^2
    \Im\bigl[(\chi_{\rm cl})^K_{11,j}(\omega,t)\bigr]_{\omega=0},
    \nonumber\\
    \langle \xi_{1,j}(t)\xi_{2,j'}(t')\rangle
    &=
    -\delta_{jj'}\delta(t-t')
    \sqrt{\frac{\Omega}{2}}\,g\Delta X_2^{\rm cl}(t)
    \nonumber\\[-2pt]
    &\qquad\times
    \Im\bigl[(\chi_{\rm cl})^K_{12,j}(\omega,t)\bigr]_{\omega=0},
    \nonumber\\
    \langle \xi_{2,j}(t)\xi_{1,j'}(t')\rangle
    &=
    -\delta_{jj'}\delta(t-t')
    \sqrt{\frac{\Omega}{2}}\,g\Delta X_2^{\rm cl}(t)
    \nonumber\\[-2pt]
    &\qquad\times
    \Im\bigl[(\chi_{\rm cl})^K_{21,j}(\omega,t)\bigr]_{\omega=0},
    \nonumber\\
    \langle \xi_{2,j}(t)\xi_{2,j'}(t')\rangle
    &=
    -\delta_{jj'}\delta(t-t')
    \frac{\Omega\Delta^2}{2}
    \bigl[X_2^{\rm cl}(t)\bigr]^2
    \nonumber\\[-2pt]
    &\qquad\times
    \Im\bigl[(\chi_{\rm cl})^K_{22,j}(\omega,t)\bigr]_{\omega=0}.
\end{align}
(The $a,b$ components of $ \big (\chi_\text{cl} \big)_{ab, j}^K$ are defined in Eq.~\eqref{eqn:chi_K_11}-\eqref{eqn:chi_K_22}.)
Within the white noise approximation, the matrix for the noise becomes symmetric:
\begin{align}
\label{eqn:xi}
    \hat{\Xi} &= 
    \begin{pmatrix}
        \langle \xi_1 \xi_1 \rangle & \langle \xi_1 \xi_2 \rangle \\
        \langle  \xi_2 \xi_1 \rangle & \langle \xi_2 \xi_2 \rangle 
    \end{pmatrix} =
    \begin{pmatrix}
        \langle \xi_1 \xi_1 \rangle & \langle \xi_1 \xi_2 \rangle \\
        \langle  \xi_1 \xi_2 \rangle & \langle \xi_2 \xi_2 \rangle 
    \end{pmatrix} \equiv
    \begin{pmatrix}
        A & C \\
        C & B 
    \end{pmatrix} \nonumber \\
    &= \frac{A + B}{2} \mathbb{1} + \sigma_z \left[ \frac{A - B}{2} \mathbb{1} + i C \sigma_y \right] .
\end{align}
By defining:
\begin{align}
    & \mathcal{E} = \sqrt{\left( \frac{A - B}{2} \right)^2 + C^2} , \\
    & \cos(2\theta) = \frac{A - B}{2\mathcal{E}} , \\
    & \sin(2\theta) = \frac{C}{\mathcal{E}} ,
\end{align}
Eq.~\eqref{eqn:xi} becomes:
\begin{align}
    \hat{\Xi}
    = e^{-i \theta \sigma_y} \left[ \frac{A + B}{2} \mathbb{1} + \mathcal{E} \sigma_z \right] e^{i \theta \sigma_y} ,
\end{align}
having defined:

\begin{align}
    U(\theta) = \cos(\theta) \mathbb{1} + i \sin(\theta) \sigma_y = 
    \begin{pmatrix}
        \cos(\theta) & \sin(\theta) \\
        -\sin(\theta) & \cos(\theta)
    \end{pmatrix} .
\end{align}
The original noise correlation matrix can be made diagonal by defining the new stochastic variables:

\begin{align}
    \begin{pmatrix}
        \eta_+ \\
        \eta_-
    \end{pmatrix}
    = U(\theta) 
    \begin{pmatrix}
        \xi_1 \\
        \xi_2
    \end{pmatrix} = 
    \begin{pmatrix}
        \cos(\theta) \xi_1 + \sin(\theta) \xi_2 \\
        -\sin(\theta) \xi_1 + \cos(\theta) \xi_2
    \end{pmatrix} ,
\end{align}
with
\begin{align}
    \label{eqn:K}
    \langle \eta_\alpha \eta_\beta \rangle = \delta_{\alpha \beta} \mathcal{K}_\alpha , \nonumber \\
    \mathcal{K}_\pm = \frac{A + B}{2} \pm \mathcal{E} .
\end{align}
By inverting the previous equation, one immediately obtains:
\begin{align}
    \begin{pmatrix}
        \xi_1 \\
        \xi_2
    \end{pmatrix} = 
    \begin{pmatrix}
        \cos(\theta) \eta_+ - \sin(\theta) \eta_- \\
        \sin(\theta) \eta_+ + \cos(\theta) \eta_-
    \end{pmatrix} .
\end{align}
Considering that:
\begin{align}
    \begin{cases}
        \cos(2\theta) = 1 - 2\sin^2(\theta) \\
        \sin(2\theta) = 2 \sin(\theta) \cos(\theta)
    \end{cases} ,
\end{align}
it is possible to get:
\begin{align} \label{eqn:sin_cos}
    \begin{cases}
        \sin(\theta) = \pm \sqrt{\frac{1 - \cos(2\theta)}{2}} \\
        \cos(\theta) = \pm \frac{\sin(2\theta)}{\sqrt{2(1 - \cos(2\theta))}}
    \end{cases} .
\end{align}
Since $\sin (2 \theta)$ is defined as the ratio of two Keldysh components of the polarization, this quantity is always positive, i.e.\ $2\theta \in [0,\pi]$, implying $\theta \in [0,\pi/2]$. As a consequence, in Eqs.~\eqref{eqn:sin_cos}, the positive signs have to be chosen.

\subsection{Equilibration procedure}
\label{app:equilibration}
\begin{figure*}
    \centering
    \includegraphics[width=1\linewidth]{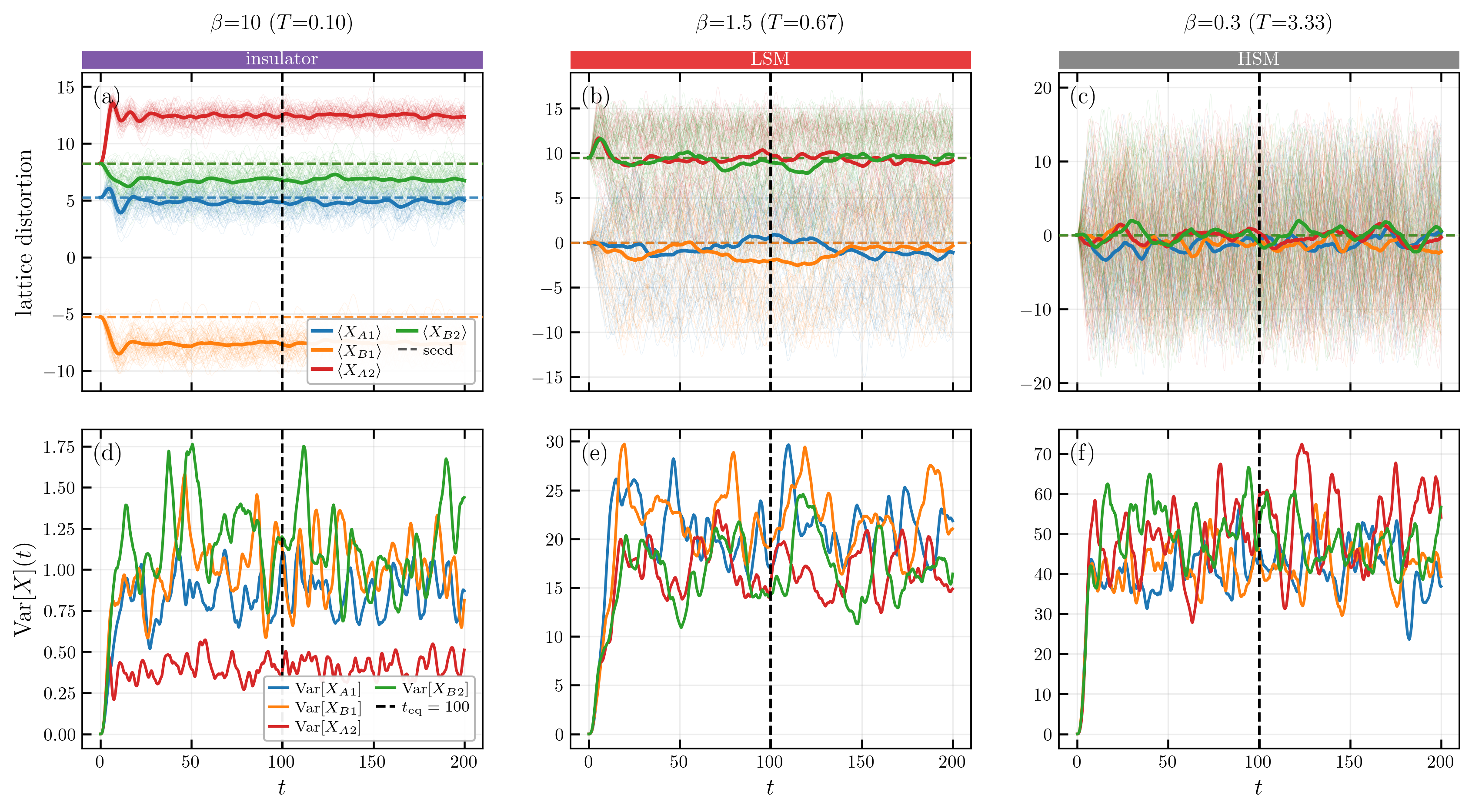}
    \caption{Equilibration of the stochastic lattice dynamics at three representative temperatures in the insulator (left), low-symmetry metal (center), and high-symmetry metal (right) regimes. (a)--(c) Time evolution of the lattice distortions on the two sublattices. Thin lines show the $64$ individual stochastic trajectories for each sublattice, while thick lines denote the corresponding ensemble averages. Horizontal dashed lines indicate the reference configurations used to initialize the dynamics. (d)--(f) Corresponding trajectory-to-trajectory variances of the lattice distortions. After an initial transient, both the ensemble averages and the variances fluctuate around stationary values. The vertical dashed lines mark $t_{\rm eq}=100$, which is chosen as the beginning of the measurement window for the equilibrium results reported in the main text.}
    \label{fig:equilibration}
\end{figure*}

We verify the equilibration of the stochastic dynamics by monitoring both the ensemble-averaged lattice distortions and their trajectory-to-trajectory variances. Figure~\ref{fig:equilibration} shows their time evolution at three representative temperatures, chosen in the I, LSM, and HSM regimes, respectively. The thin lines in the upper panels show the individual stochastic trajectories, while the thick lines denote the corresponding ensemble averages. The lower panels show the variances across the ensemble.

Starting from the reference configurations described in the main text, the stochastic dynamics initially develops lattice fluctuations and subsequently reaches a stationary regime. In particular, after the initial transient the trajectory-to-trajectory variances fluctuate around stationary values at all three representative temperatures. We conservatively choose $t_{\rm eq}=100$ as the beginning of the measurement window, as indicated by the vertical dashed lines in Fig.~\ref{fig:equilibration}. Equilibrium observables reported in the main text are evaluated only for $t\geq t_{\rm eq}$.

\end{document}